\documentclass[reprint,superscriptaddress,amsmath,amssymb,aps,pre,longbibliography]{revtex4-2}

\usepackage{graphicx}

\usepackage{bm}
\usepackage{bbm}

\usepackage[dvipsnames]{xcolor}
\definecolor{seaborn_deep_blue}{RGB}{76, 114, 176}

\usepackage{hyperref}
\hypersetup{pdftitle={},
            pdfauthor={},
            pdfsubject={},
            colorlinks=true,
            allcolors=seaborn_deep_blue}

\usepackage{pdfpages}
\usepackage{pgffor}
\makeatletter
\AtBeginDocument{\let\LS@rot\@undefined}
\makeatother
\def\supplementfilename{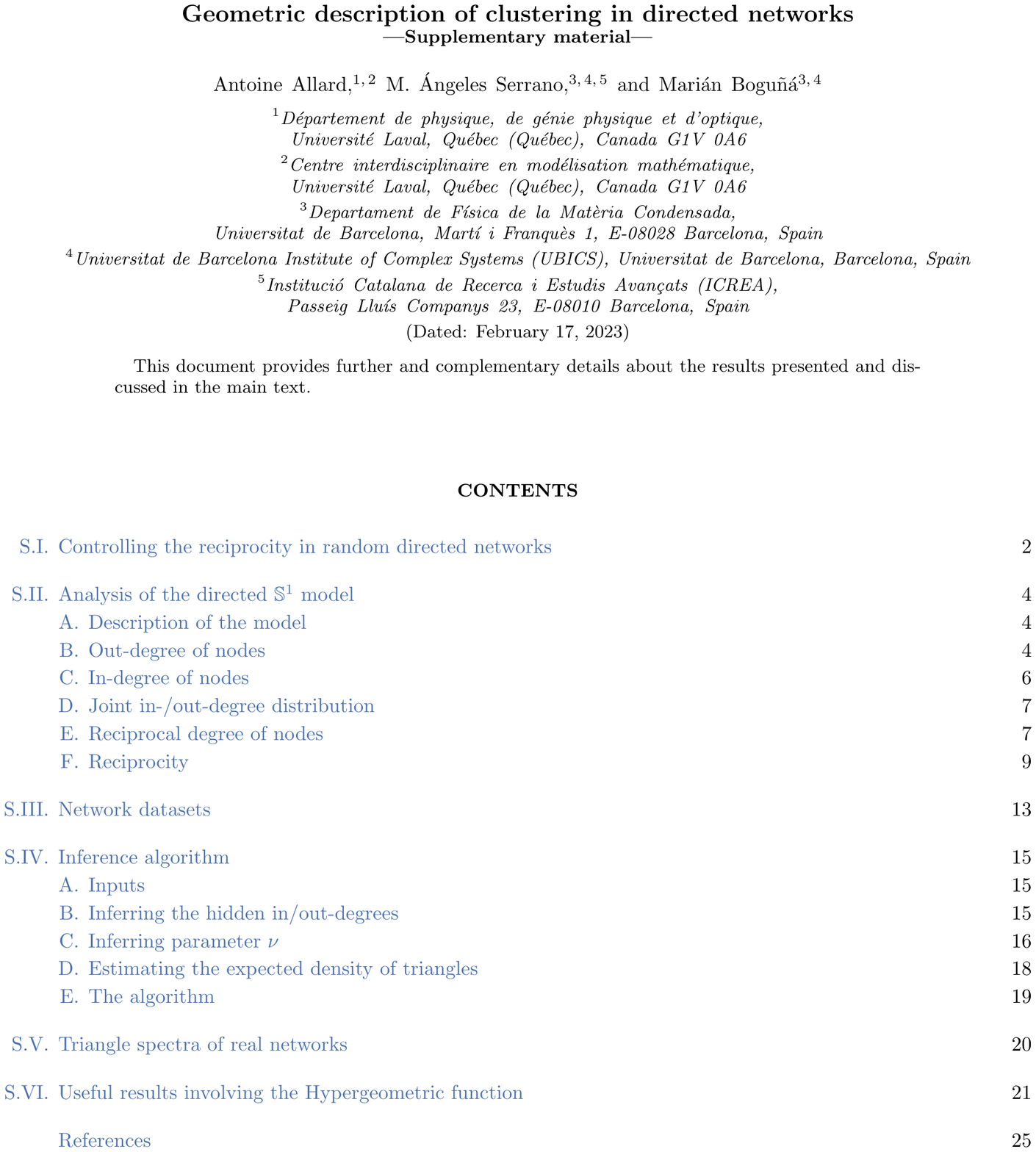}
\pdfximage{\supplementfilename}
\def\numbersupplementpages{\the\pdflastximagepages}
\newif\ifarXiv
\arXivtrue

\newcommand{\tagin}{{-}}
\newcommand{\tagout}{{+}}
\newcommand{\tagrec}{{\leftrightarrow}}

\newcommand{\kin}{\kappa^\tagin}
\newcommand{\kout}{\kappa^\tagout}

\newcommand{\din}{k^\tagin}
\newcommand{\dout}{k^\tagout}
\newcommand{\drec}{k^\tagrec}

\newcommand{\Expected}[1]{\left\langle #1 \right\rangle}
\newcommand{\ExpectedCond}[2]{\left\langle #1 \middle| #2 \right\rangle}

\begin{document}
\title{Geometric description of clustering in directed networks}
\author{Antoine Allard}%
\affiliation{D\'epartement de physique, de g\'enie physique et d'optique, Universit\'e Laval, Qu\'ebec (Qu\'ebec), Canada G1V 0A6}%
\affiliation{Centre interdisciplinaire en mod\'elisation math\'ematique, Universit\'e Laval, Qu\'ebec (Qu\'ebec), Canada G1V 0A6}%
\author{M. \'Angeles Serrano}%
\affiliation{Departament de F\'isica de la Mat\`eria Condensada, Universitat de Barcelona, Mart\'i i Franqu\`es 1, E-08028 Barcelona, Spain}%
\affiliation{Universitat de Barcelona Institute of Complex Systems (UBICS), Universitat de Barcelona, Barcelona, Spain}%
\affiliation{Instituci\'o Catalana de Recerca i Estudis Avan\c{c}ats (ICREA), Passeig Llu\'is Companys 23, E-08010 Barcelona, Spain}%
\author{Marián Bogu\~n\'a}%
\affiliation{Departament de F\'isica de la Mat\`eria Condensada, Universitat de Barcelona, Mart\'i i Franqu\`es 1, E-08028 Barcelona, Spain}%
\affiliation{Universitat de Barcelona Institute of Complex Systems (UBICS), Universitat de Barcelona, Barcelona, Spain}%
\date{\today}
\begin{abstract}
  First principle network models are crucial to make sense of the intricate topology of real complex networks.  While modeling efforts have been quite successful in undirected networks, generative models for networks with asymmetric interactions are still not well developed and are unable to reproduce several basic topological properties.  This is particularly disconcerting considering that real directed networks are the norm rather than the exception in many natural and human-made complex systems.  In this paper, we fill this gap and show how the network geometry paradigm can be elegantly extended to the case of directed networks.  We define a maximum entropy ensemble of geometric (directed) random graphs with a given sequence of in- and out-degrees.  Beyond these local properties, the ensemble requires only two additional parameters to fix the level of reciprocity and the seven possible types of 3-node cycles in directed networks.  A systematic comparison with several representative empirical datasets shows that fixing the level of reciprocity alongside the coupling with an underlying geometry is able to reproduce the wide diversity of clustering patterns observed in real complex directed networks.
\end{abstract}
\maketitle
%
%
%
%
\section{Introduction}
%
The network geometry paradigm is a comprehensive framework that successfully explains the topology, multiscale organization, and navigability of real complex networks~\cite{boguna2021network}.  Consisting of a handful of simple models, this framework has been shown to accurately model several features observed in static, growing, weighted, or multilayer networks~\cite{serrano2008selfsimilarity, garcia-perez2018multiscale, zheng2021scaling, boguna2020small, papadopoulos2012popularity, allard2017geometric, kleineberg2016hidden}.  The hallmark of network geometry is how it naturally reproduces the clustering patterns observed in real complex networks, one of their most fundamental properties~\cite{newman2018networks}.  Clustering is indeed notoriously difficult to model because triangles imply three-node interactions, and most existing approaches must rely on approximations such as an underlying tree-like organization~\cite{allard2015general, gleeson2009analytical, newman2003properties, karrer2010random, miller2009percolation, battiston2020networks}, give up sparsity~\cite{lee2019review}, or turn to numerical simulations~\cite{orsini2015quantifying, serrano2005tuning, volz2004random}.

Network geometry overcomes this difficulty by assuming that nodes are embedded in a metric space, and that the probability $p_{ij}$ that a link exists between nodes $i$ and $j$ is a decreasing function of the distance between them.  Non-fortuitous clustering---clustering that does not occur by sheer luck---can therefore be seen as the topological counterpart of the triangle inequality of the metric space: if nodes $j$ and $l$ are both close to node $i$, then they must also be close to each other.  Hence a triangle composed of nodes $i$, $j$ and $l$ is likely, even in the limit of very large networks.  In fact, network geometry interprets the clustering coefficient as a measure of the coupling between the topology of the network and an underlying latent metric space.

However, to date, network geometry has only been fully developed for complex networks with \textit{symmetric} interactions, weighted or unweighted.  Yet, a large number of real complex networked systems contain a mixture of symmetric and asymmetric interactions (e.g.~connectomes, food webs, and communication networks)~\cite{asllani2018structure, johnson2020digraphs, newman2018networks}.  In addition to the ubiquity of asymmetry, such systems are relevant because they represent processes out of equilibrium where detailed balance is not fulfilled.  These systems are also typically non-normal~\cite{asllani2018structure} and display trophic coherence~\cite{levine1980several} (or lack thereof) both of which have a drastic impact of their dynamics, an impact that cannot be foreseen if the directionality of the interactions were simply neglected~\cite{asllani2018structure, duan2022network, johnson2014trophic, johnson2017looplessness, johnson2020digraphs, klaise2016neurons, nicolaou2020nonnormality, qu2014nonconsensus, shao2009dynamic}.  Although extensions have recently been explored~\cite{michel2019directed, wolf2019edge, wu2020asymmetric, kovacs2022modelindependent, peralta-martinez2022directed}, the apparent contradiction between the symmetry of metric distances and asymmetric interactions has kept this important class of systems out of the reach of the network geometry framework.

\begin{figure}
  \includegraphics[width=0.9\linewidth]{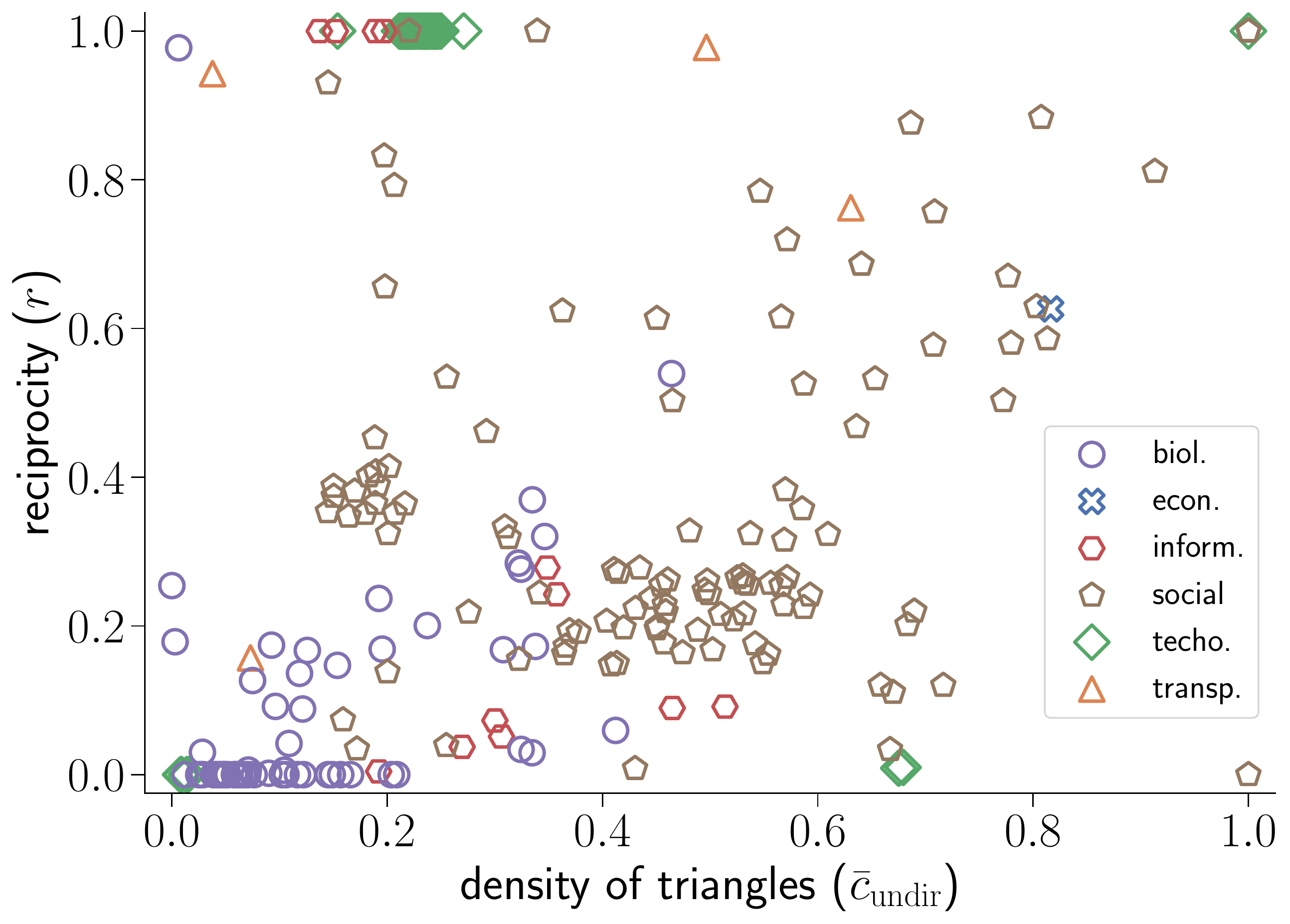}%
  \caption{%
    \textbf{Reciprocity in real directed networks.}
    Reciprocity vs. density of triangles in 292 real directed networks.  The reciprocity is defined as $r= L^\tagrec/L$, where $L^\tagrec$ is the number of reciprocal links and $L$ is the number of links.  The density of triangles is computed as the average local clustering coefficient of the undirected projection of the original directed network ($\bar{c}_\mathrm{undir}$; see Methods). Details about the network datasets are provided in Methods.
  }
  \label{fig:reciprocity_real}
\end{figure}

In this paper, we propose a simple solution to this impasse.  By rethinking the relationship between distance and connection, we introduce a general and versatile adaptation of the framework of network geometry that reconciles the intrinsic symmetry of metric distances with asymmetric interactions between nodes in directed networks.  Our model is able to reproduce both the joint distribution of in-degrees and out-degrees as well as the number of the different classes of triangles with only one additional parameter that tunes the level of reciprocity---the propensity for the two different directed links to exist between the same pair of nodes---, a fundamental property of real directed networks~\cite{wasserman1994social, garlaschelli2004patterns}, see Fig.~\ref{fig:reciprocity_real}. It is also amenable to several analytical and semi-analytical calculations.  In addition, our methodology can also be used to control the level of reciprocity in any non-geometric model as long as it defines pairwise connection probabilities.

\begin{figure*}[t]
  \centering
  \includegraphics[width=0.9\linewidth]{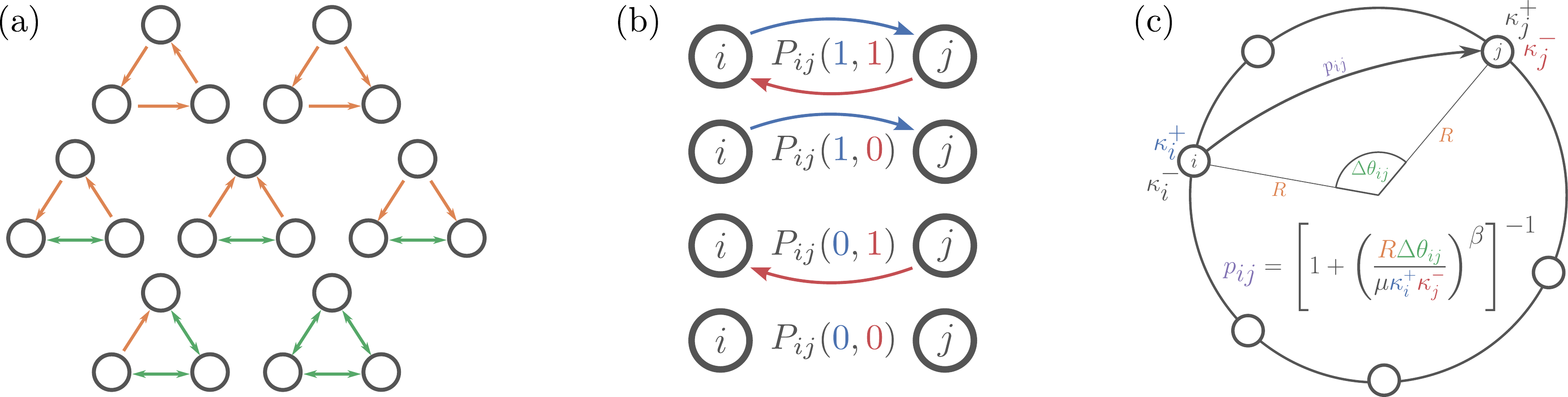}%
  \caption{%
    \textbf{Illustrations of the concepts behind the modeling framework.}
    (a) The 7 configurations of triangles in directed networks~\cite{holland1976local, ahnert2008clustering}.
    (b) The joint probabilities $P_{ij}(a_{ij},a_{ji})$ used in the general framework controlling reciprocity in random directed networks.
    (c) The geometric directed soft configuration model where $p_{ij}$ stands for the probability of connection $P(a_{ij} = 1|\kout_i, \kin_j, \Delta\theta_{ij})$ of Eq.~\eqref{eq:connection_probability}.
  }
  \label{fig:cartoon}
\end{figure*}

Most importantly, we use our approach to show that the even more complex patterns of clustering in directed networks---quantified by the relative occurrence of the 7 triangle configurations possible with directed links, or triangle spectrum, see Fig.~\ref{fig:cartoon}(a)---are in fact a byproduct of the joint distribution of in-degree and out-degree, of reciprocity and of the triangle inequality in the underlying metric space.  Our contribution offers a rigorous path to extend network geometry to directed networks, thus allowing this powerful approach to be used to study several real complex systems where asymmetric interactions are important, like the brain, food webs, information networks, and human interactions.

\section{Results}
%
\subsection{Reciprocity in random directed networks}\label{sec:reciprocity}
%
We first introduce a general framework to control the level of reciprocity in any random directed network models with pairwise connection probabilities.  Let $p_{ij}$ be the probability for a directed link to exist from node $i$ to node $j$, and $N$ be the number of nodes.  The assumption that interactions are pairwise implies that the existence of links between two different pairs of nodes, $i,j$ and $k,l$, are statistically independent events.  If this condition also applies to the two possible links between the same pair of nodes $i,j$, then the probability to have a reciprocal link is simply $p_{ij}p_{ji}$. Therefore, doing so generates a certain level of reciprocity in the network, although it is not possible to tune it.

To gain control over reciprocity, we must relax the assumption of independence within the same pair of nodes.  Thus, similarly to the seminal dyad independence model~\cite{holland1981exponential}, our framework focuses on the four ways two nodes may or may not be connected, see Fig.~\ref{fig:cartoon}(b).  We define the joint probabilities $P_{ij}(a_{ij}, a_{ji})$ with $1\leq i<j\leq N$ and where $a_{ij}$ is 1 if there is a directed link from node $i$ to node $j$, and 0 otherwise.  For our framework to be coherent with the model defining the pairwise connection probabilities, we impose that the joint probability $P_{ij}(a_{ij}, a_{ji})$ preserves the marginal connection probabilities so that
\begin{subequations}
\label{eq:marginal_conn_prob}
\begin{align}
  P_{ij}(1,0) + P_{ij}(1,1) & = p_{ij} \\
  P_{ij}(0,1) + P_{ij}(1,1) & = p_{ji} \ ,
\end{align}
\end{subequations}
and we assume that they are normalized
\begin{align} \label{eq:normalization_conn_joint_prob}
  \sum_{a_{ij}=0}^{1} \sum_{a_{ji}=0}^{1} P_{ij}(a_{ij}, a_{ji}) = 1
\end{align}
for every pair $(i,j)$.  Equations~\eqref{eq:marginal_conn_prob} and~\eqref{eq:normalization_conn_joint_prob} leave one of the four probabilities $P_{ij}(a_{ij}, a_{ji})$ undefined, giving the model an extra degree of freedom to fix the reciprocity of the network. This can be done by considering the correlation coefficient
\begin{subequations}
\label{eq:rho_ij}
\begin{align}
  \rho_{ij} & = \frac{\Expected{a_{ij}a_{ji}} - \Expected{a_{ij}} \Expected{a_{ji}}}%
                  {\sqrt{\left(\Expected{a_{ij}^2} - \Expected{a_{ij}}^2\right)\left(\Expected{a_{ji}^2} - \Expected{a_{ji}}^2\right)}}
                  \label{eq:rho_ij_1} \\
            & = \frac{P_{ij}(1,1) - p_{ij}p_{ji}}{\sqrt{p_{ij}(1 - p_{ij})p_{ji}(1 - p_{ji})}} \ .
                  \label{eq:rho_ij_2}
\end{align}%
\end{subequations}%
where $\Expected{\cdot}$ corresponds to an average over the network ensemble defined by the joint probabilities.  Note that, because $P_{ij}(1,1)\in [0,1]$, Eq.~\eqref{eq:rho_ij} is not guaranteed to be bounded between -1 and 1.  Enforcing these bounds yields an expression for $P_{ij}(1,1)$ in terms of $p_{ij}$, $p_{ji}$ and a parameter $\nu \in [-1,1]$ controlling the level of reciprocity between nodes $i$ and $j$
\begin{align} \label{eq:Pij11}
  P_{ij}(1,1) =
    \begin{cases}
      (1 + \nu) p_{ij} p_{ji}\\ \qquad + \nu (1 - p_{ij} - p_{ji}) H(p_{ij} + p_{ji} - 1)\\  \hspace{0.45\linewidth} \text{for } -1 \leq \nu \leq 0 \\
      (1 - \nu) p_{ij} p_{ji}          + \nu \min\Big\{p_{ij},p_{ji}\Big\}\\                 \hspace{0.45\linewidth} \text{for }  0 \leq \nu \leq 1 \ ,
    \end{cases}
\end{align}
where $H(\cdot)$ is the Heaviside step function (a detailed derivation is provided in the Supplementary Material).  For instance, the cases $\nu=1, 0, -1$ correspond, respectively, to the highest level of reciprocity that is structurally possible, random reciprocity (i.e. directed links exist in both directions with probability $p_{ij}p_{ji}$) and \textit{anti}-reciprocity meaning the minimum level of reciprocity achievable given the joint probabilities.  Note that fully reciprocal networks ($r = 1$) are only possible when $\nu = 1$ and $p_{ij} = p_{ji}$ for every pair of nodes $i$ and $j$.

Alongside Eqs.~\eqref{eq:marginal_conn_prob}~and~\eqref{eq:normalization_conn_joint_prob}, Eq.~\eqref{eq:Pij11} fully defines the four joint probabilities $P_{ij}(a_{ij}, a_{ji})$ prescribing how nodes $i$ and $j$ are connected, and thus the level of reciprocity in the network ensemble.  The latter can be made explicit by computing the expected reciprocity~\cite{garlaschelli2004patterns}
\begin{align}
  \Expected{r}
    = \Expected{\frac{L^{\leftrightarrow}}{L}}
    \approx \frac{\Expected{L^{\leftrightarrow}}}{\Expected{L}}
    = \frac{\Expected{\drec}}{\Expected{\dout}} \ ,
  \label{eq:expected_reciprocity}
\end{align}
where $L$ is the number of links, $L^{\leftrightarrow}$ is the number of links that are reciprocated (i.e. a directed link that has another link running in the opposite direction), and where
\begin{align}
  \Expected{\dout}
    = \Expected{\din}
    = \frac{1}{N} \sum_{i=1}^N \sum_{\substack{j = 1\\j \neq i}}^N p_{ij} \ ,
\end{align}
is the expected degree (in or out) and
\begin{align}
  \label{eq:average_reciprocal_degree}
  \Expected{\drec}
    = \frac{2}{N} \sum_{i=1}^N \sum_{j = i + 1}^N P_{ij}(1,1)
\end{align}
is the expected reciprocated degree.

\begin{figure}
  \includegraphics[width=0.9\linewidth]{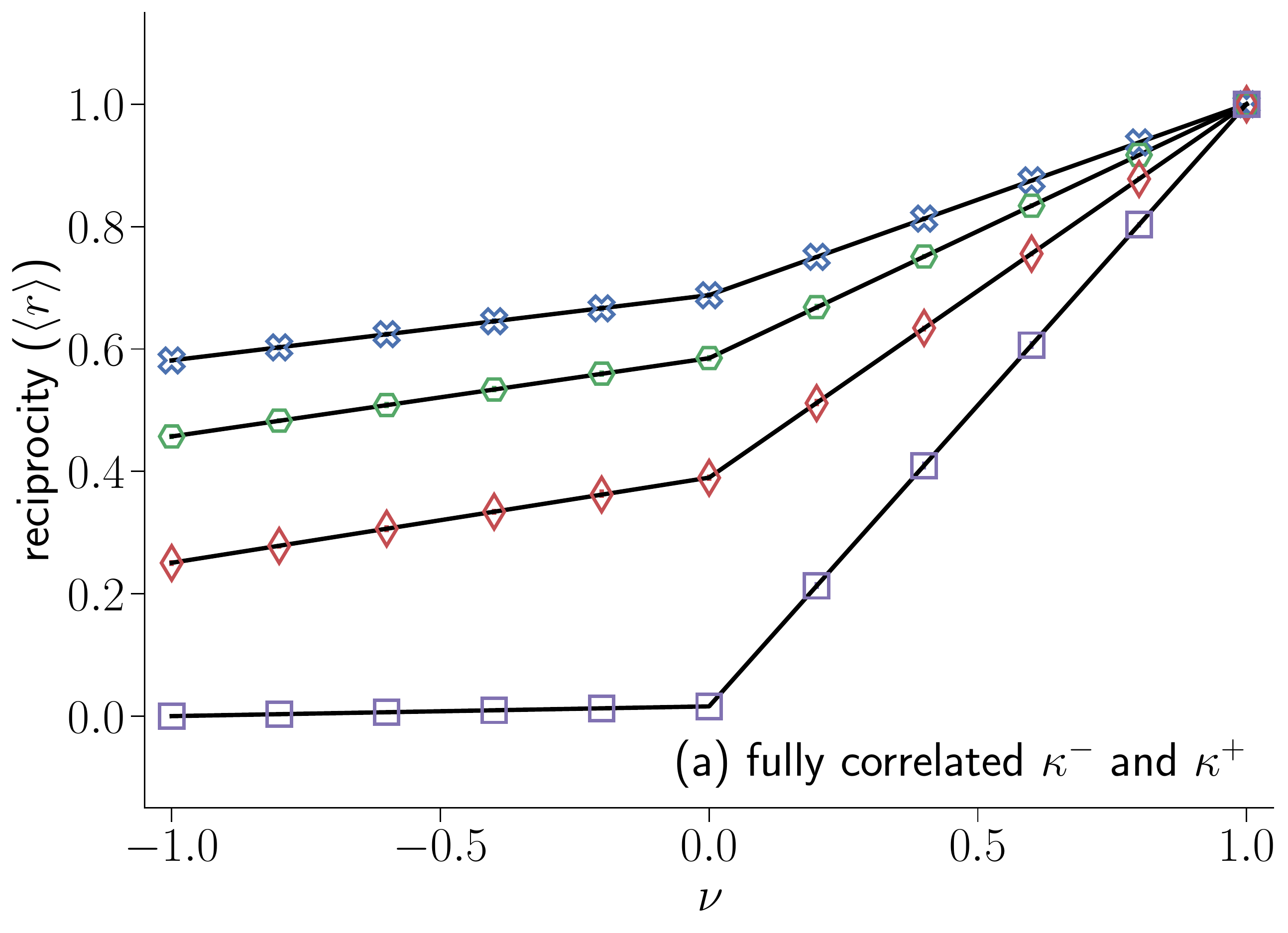}\\%
  \includegraphics[width=0.9\linewidth]{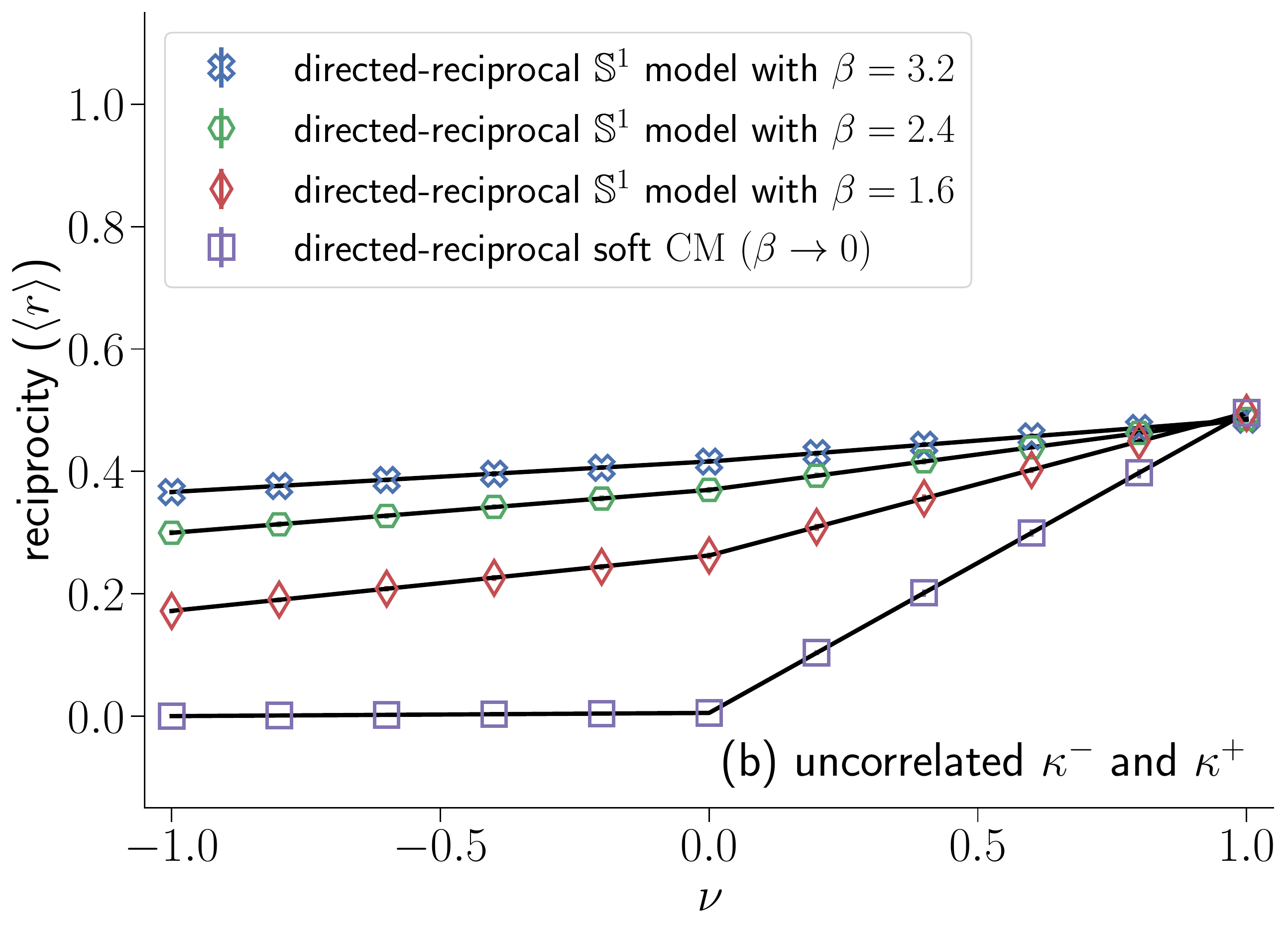}%
  \caption{%
    \textbf{Validation of the general framework controlling reciprocity.}
    We consider both the directed-reciprocal soft configuration model (see Methods) and the directed-reciprocal $\mathbb{S}^1$ model (see main text).
    Each symbol shows $\Expected{r}$ estimated from 100 random synthetic networks composed of $N=2500$ nodes.  Solid lines show the predictions of Eqs.~\eqref{eq:Pij11}--\eqref{eq:average_reciprocal_degree}.  Error bars show the estimated 95\% confidence interval (almost always smaller than the width of the solid lines).  To highlight the dependency of $\Expected{r}$ on $\beta$ and on the correlation between $\kin$ and $\kout$, we drew a sequence $\{\kin_i\}_{i=1,\ldots,N}$ from the pdf $\rho(\kappa) \propto \kappa^{-2.5}$ with $5 < \kappa < 100$ and a sequence $\{\theta_i\}_{i=1,\ldots,N}$ from the pdf $\varphi(\theta) = \frac{1}{2\pi}$.  All symbols and lines were obtained using these two sequences.
    (a) We set $\kout_i = \kin_i$ for $i=1,\ldots,N$ to fully correlate $\kin$ and $\kout$.
    (b) We shuffled the sequence $\{\kout\}_{i=1,\ldots,N}$ used in (a) to decorrelate $\kin$ and $\kout$.
  }
  \label{fig:reciprocity_synth}
\end{figure}%

\subsection{The directed \texorpdfstring{$\mathbb{S}^1$}{S1} model}
%
Although it allows for the control of the level of reciprocity, the framework introduced above does not by itself generate networks with clustering levels beyond fortuitous clustering due to three nodes being connected by chance.  We now introduce a generalization of the $\mathbb{S}^1$ model~\cite{serrano2008selfsimilarity} to directed networks (the \textit{directed} $\mathbb{S}^1$ model) which generates networks with nontrivial levels of clustering, even in the limit $N \to \infty$. However, note that this extension to directed networks of the geometric soft configuration model generates reciprocal links only by chance.  In the next subsection, we will combine the two approaches to propose the definitive formulation of the \textit{directed-reciprocal} $\mathbb{S}^1$ model

The ensemble of random directed networks defined by the directed $\mathbb{S}^1$ model consists in $N$ nodes positioned on a circle of radius $R = N / 2 \pi$ (thus setting the density of nodes to 1 without loss of generality).  Each node $i$ is independently and identically assigned an angular position $\theta_i$ and a pair of \textit{hidden} degrees $\kappa_i^\tagin$ and $\kappa_i^\tagout$ which, as shown below, are related to their in- and out-degree, respectively.  The angular positions are scattered on the circle according to the uniform probability density function (pdf) $\varphi(\theta) = \frac{1}{2\pi}$, although other densities---for instance to include community structure~\cite{garcia-perez2018soft, muscoloni2018nonuniform,desy2022dimension}---could be considered. The hidden degrees are also assigned randomly according to the joint pdf $\rho(\kin, \kout)$, whose only constraint is on its two first moments: $\Expected{\kin} = \Expected{\kout} \equiv \Expected{\kappa}$.

A directed link exists from node $i$ to node $j$ with probability
\begin{align} \label{eq:connection_probability}
  P(a_{ij} = 1|\kout_i, \kin_j, \Delta\theta_{ij}) = \frac{1}{1 + \chi_{ij}^\beta}
\end{align}
with
\begin{align}
  \chi_{ij} = \frac{R\Delta\theta_{ij}}{\mu \kout_i \kin_j} = \frac{N\Delta\theta_{ij}}{2 \pi \mu \kout_i \kin_j} \ ,
\end{align}
where $\Delta \theta_{ij} = \Delta \theta_{ji} = \pi - | \pi - |\theta_i - \theta_j||$ is the minimal angular distance between nodes $i$ and $j$, and where $\mu=\frac{\beta}{2 \pi \Expected{\kappa}} \sin\left(\frac{\pi}{\beta}\right)$ and $\beta > 1$ is a parameter of the model that controls clustering, as we explain below.  Figure~\ref{fig:cartoon}(c) provides an illustration of the model.

The choice of Eq.~\eqref{eq:connection_probability} has two advantages.  First, it casts the ensemble of random networks generated by the model into a hyper-grand-canonical ensemble, which is a prime candidate to be the unbiased maximum entropy spatial network models for sparse heterogeneous small worlds with nonzero clustering~\cite{boguna2020small}.  Second, fixing the hidden degrees $\kin$ and $\kout$ allows specifying the expected in- and out-degree of each node, and thus the expected joint in- and out-degree distribution.  As shown in the Supplementary Material, the expected in- and out-degrees of nodes with hidden variables $\kin_i, \kout_i$ are simply given by
\begin{align}
 \ExpectedCond{\din_i}{\kin_i} \simeq \kin_i%
 \; \;  \text{and} \; \;
 \ExpectedCond{\dout_i}{\kout_i} \simeq \kout_i \ .
\end{align}
As shown in Methods, the generalization of the $\mathbb{S}^1$ model presented here can be seen as the geometric extension of the directed soft configuration model which, unlike its nongeometric counterpart, has a nonvanishing clustering in the limit $N \to \infty$ (due to the triangle inequality of its embedding space).  As in the undirected $\mathbb{S}^1$ model, clustering in this generalization is tuned using the parameter $\beta$; the limit $\beta \to \infty$ yielding the highest density of triangles, while clustering goes to zero when $\beta=1$.  The detailed derivation of these results as well as their validation using numerical simulations are provided in the Supplementary Material.

\begin{figure*}[t]
  \hfill%
  \raisebox{2.45ex}{\includegraphics[width=0.332\linewidth]{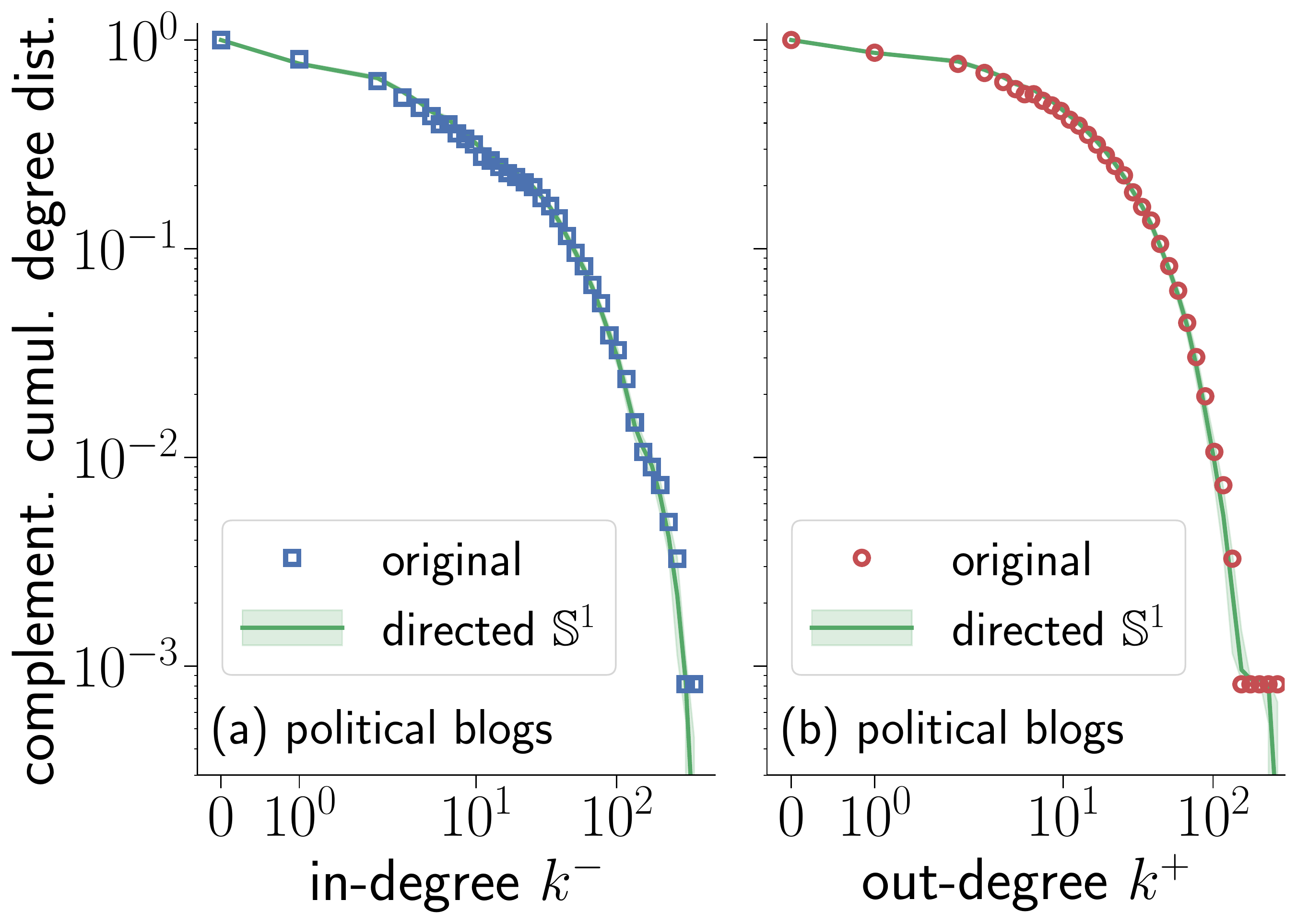}}%
  \raisebox{2.45ex}{\includegraphics[width=0.332\linewidth]{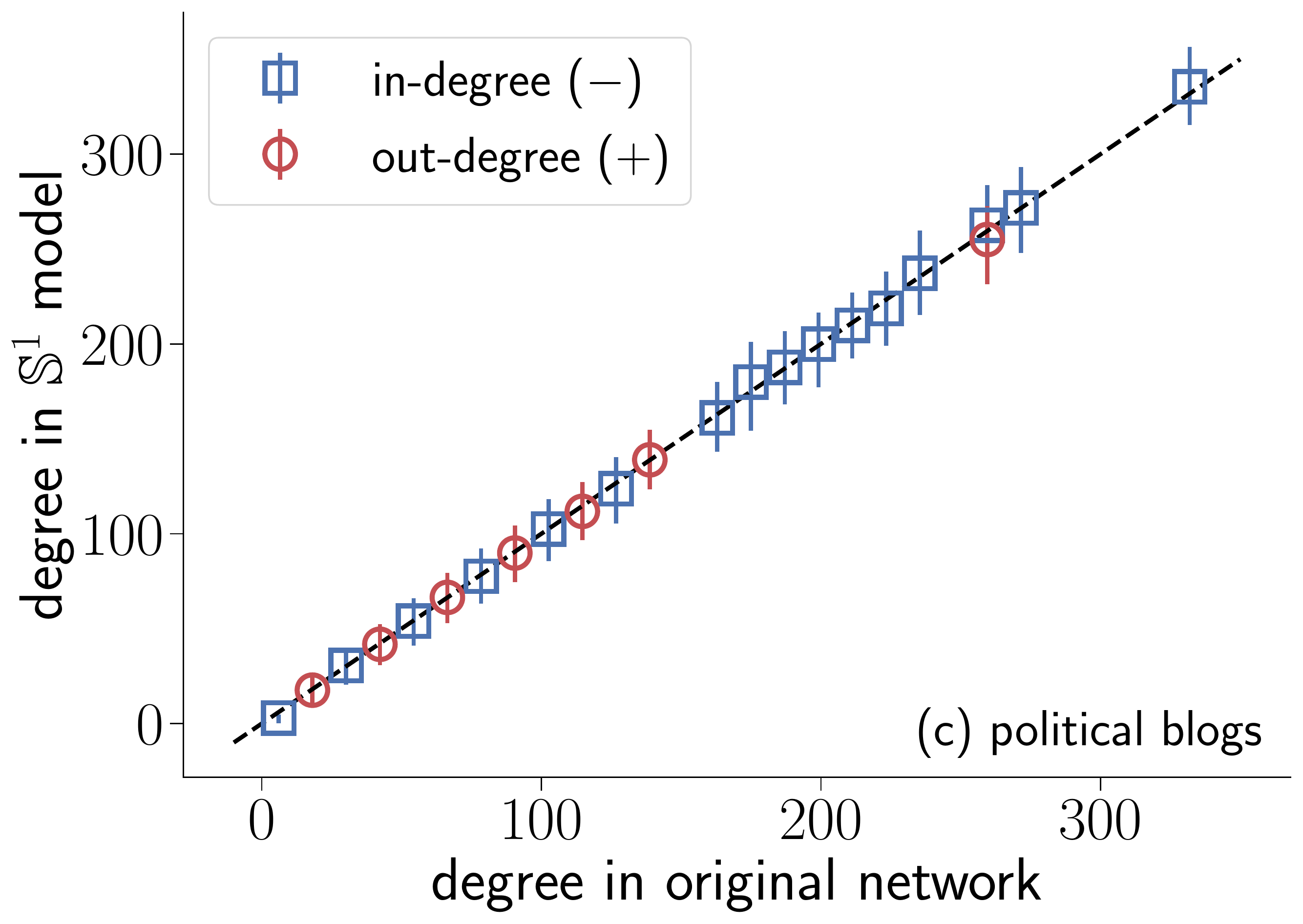}}%
  \includegraphics[width=0.332\linewidth]{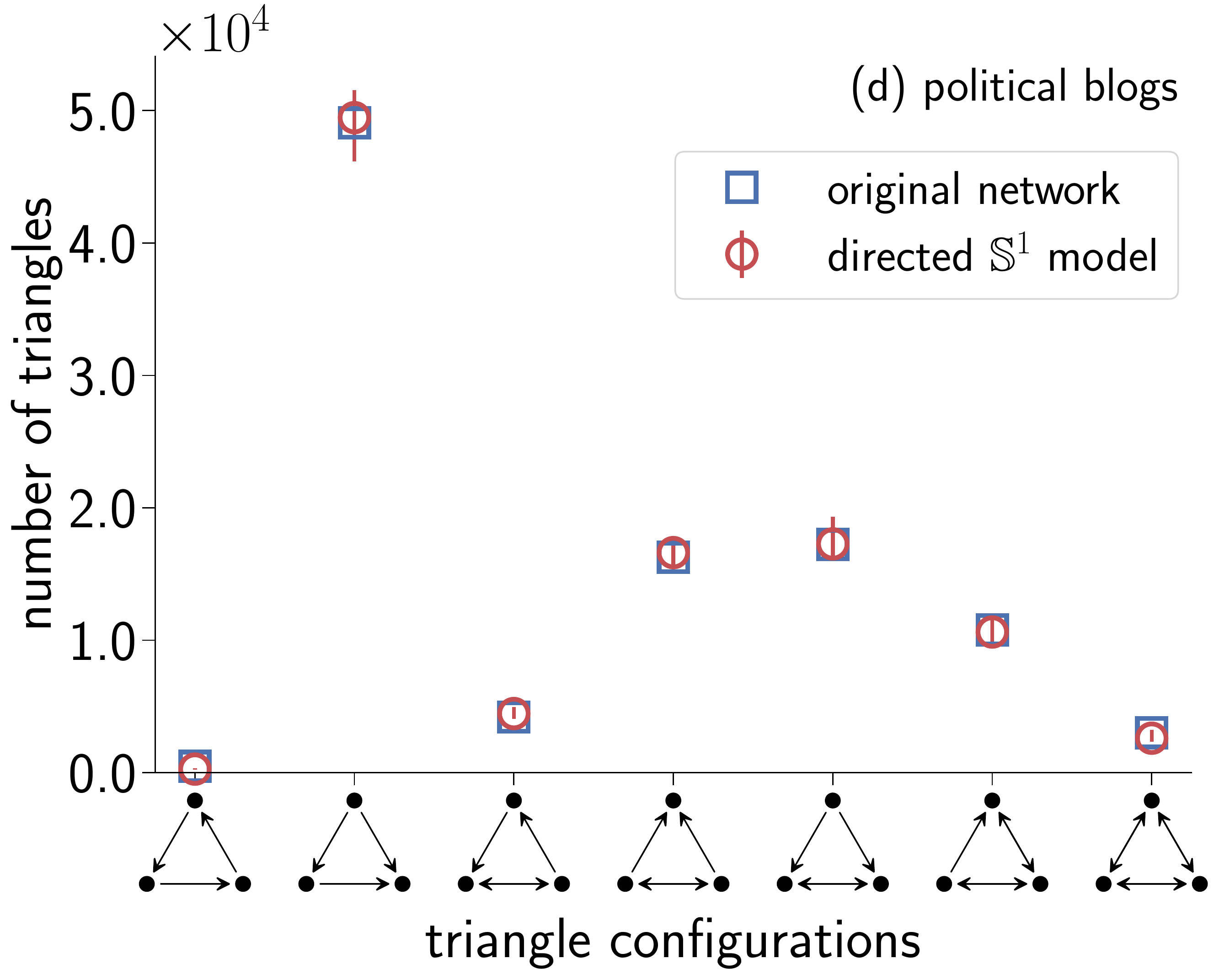}\\%
  \hfill%
  \includegraphics[width=0.245\linewidth]{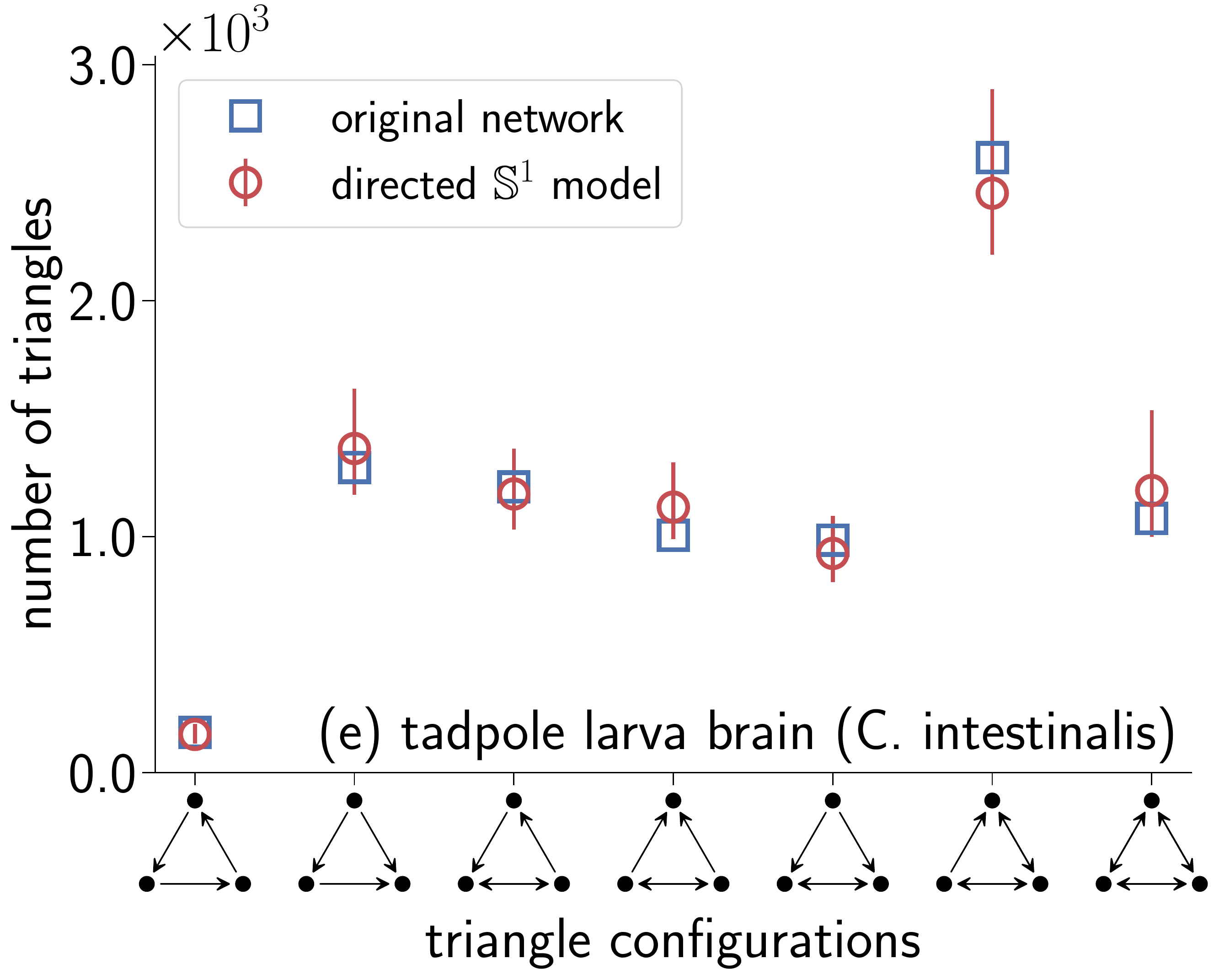}%
  \includegraphics[width=0.245\linewidth]{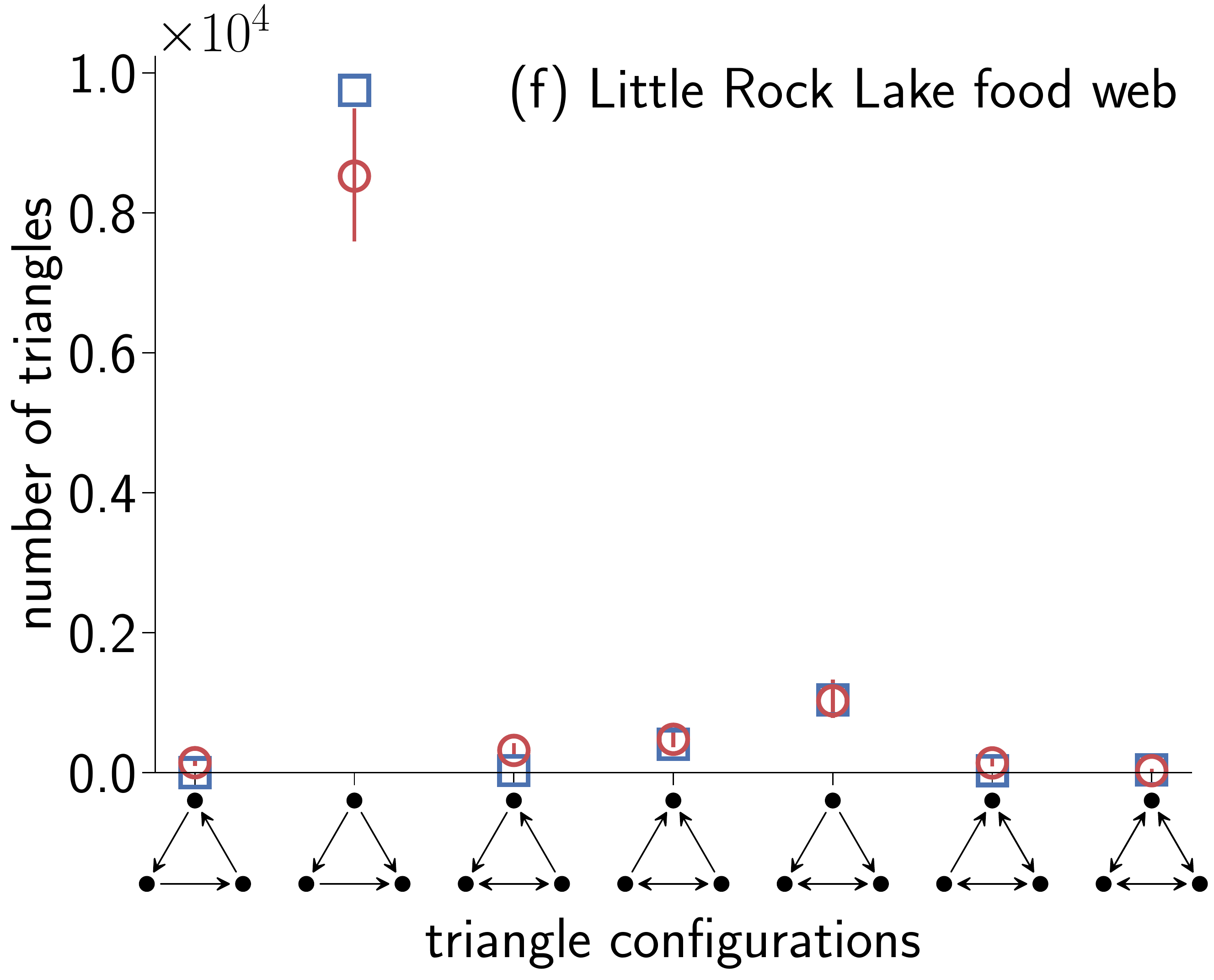}%
  \includegraphics[width=0.245\linewidth]{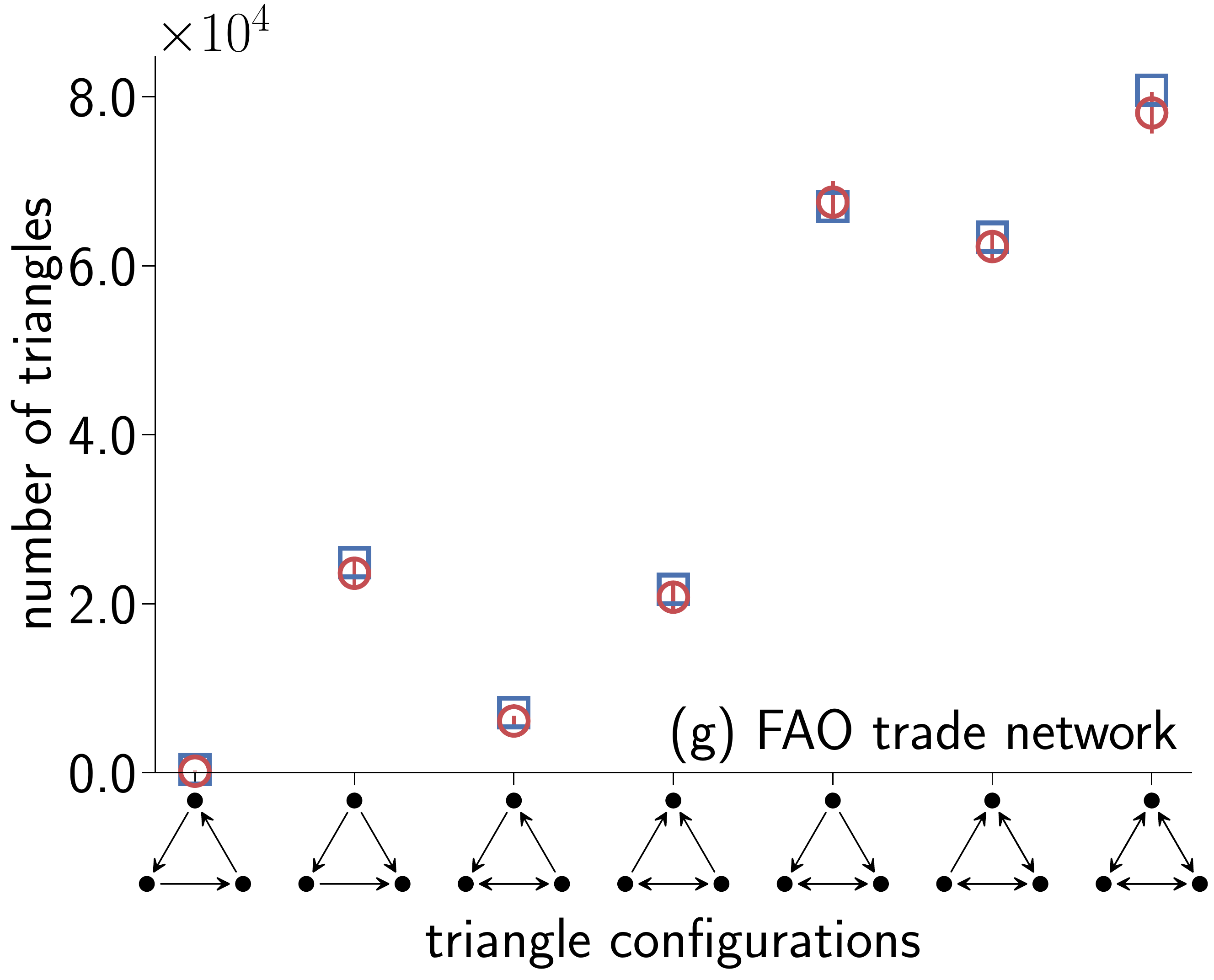}%
  \includegraphics[width=0.245\linewidth]{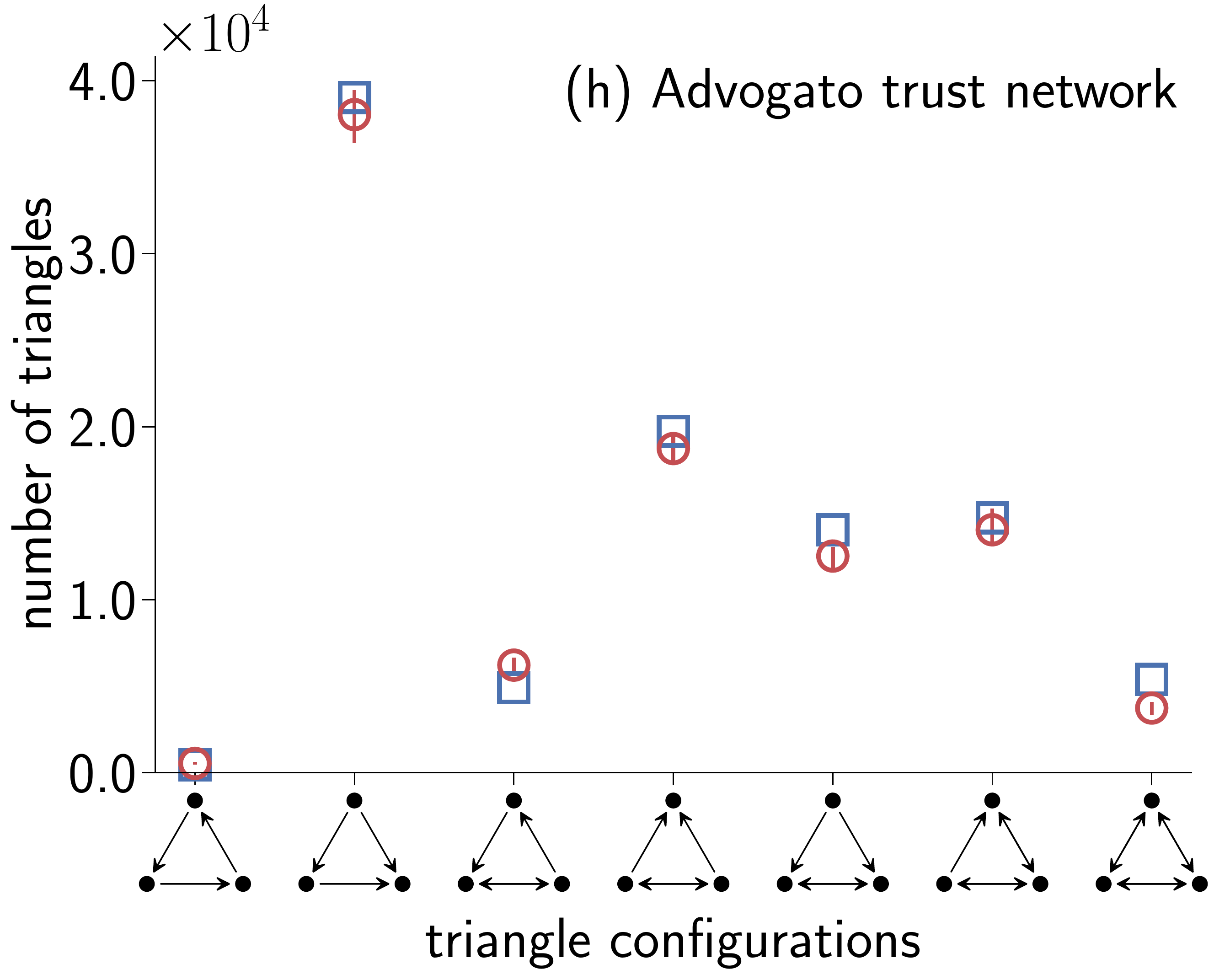}\\%
  \hfill%
  \includegraphics[width=0.245\linewidth]{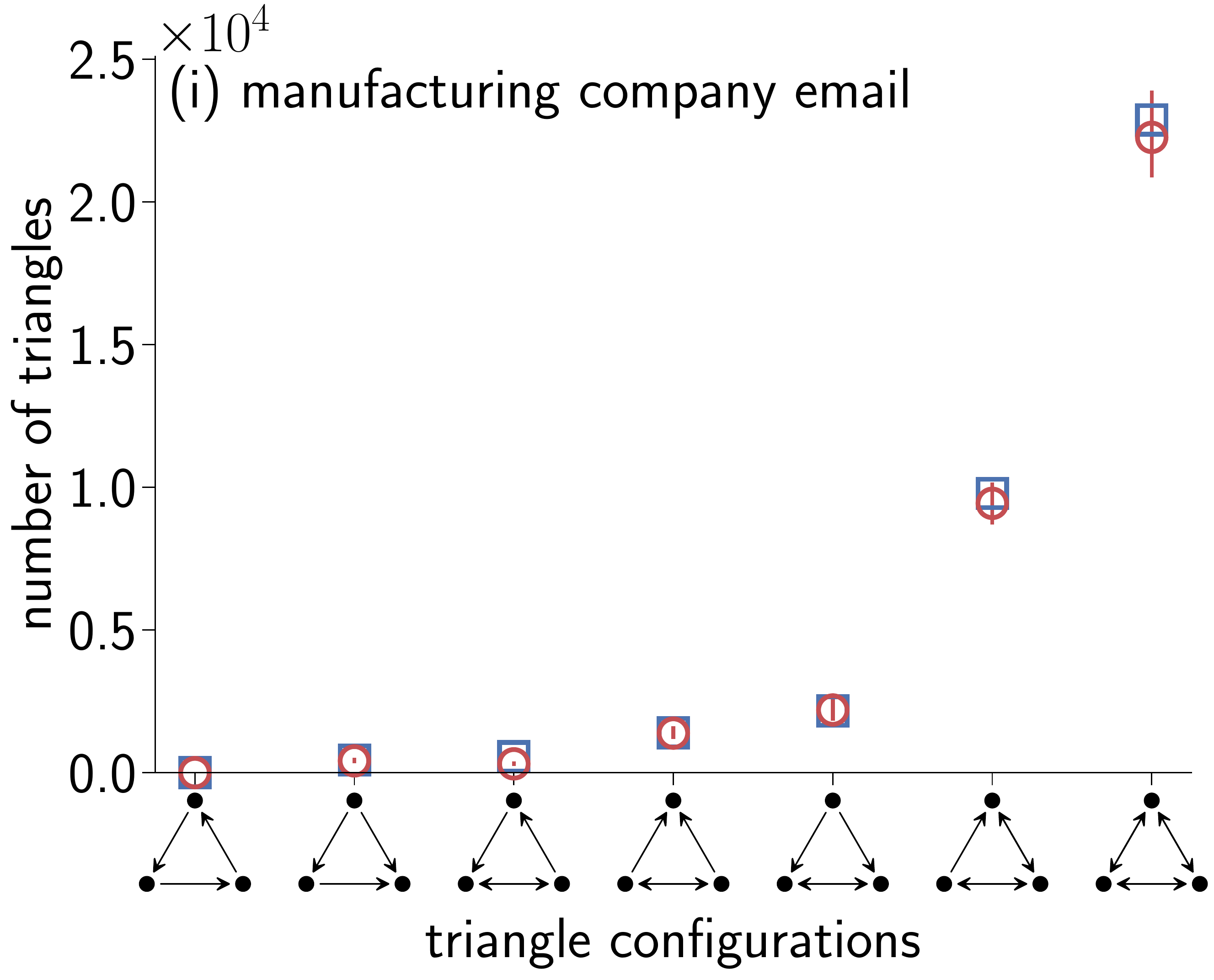}%
  \includegraphics[width=0.245\linewidth]{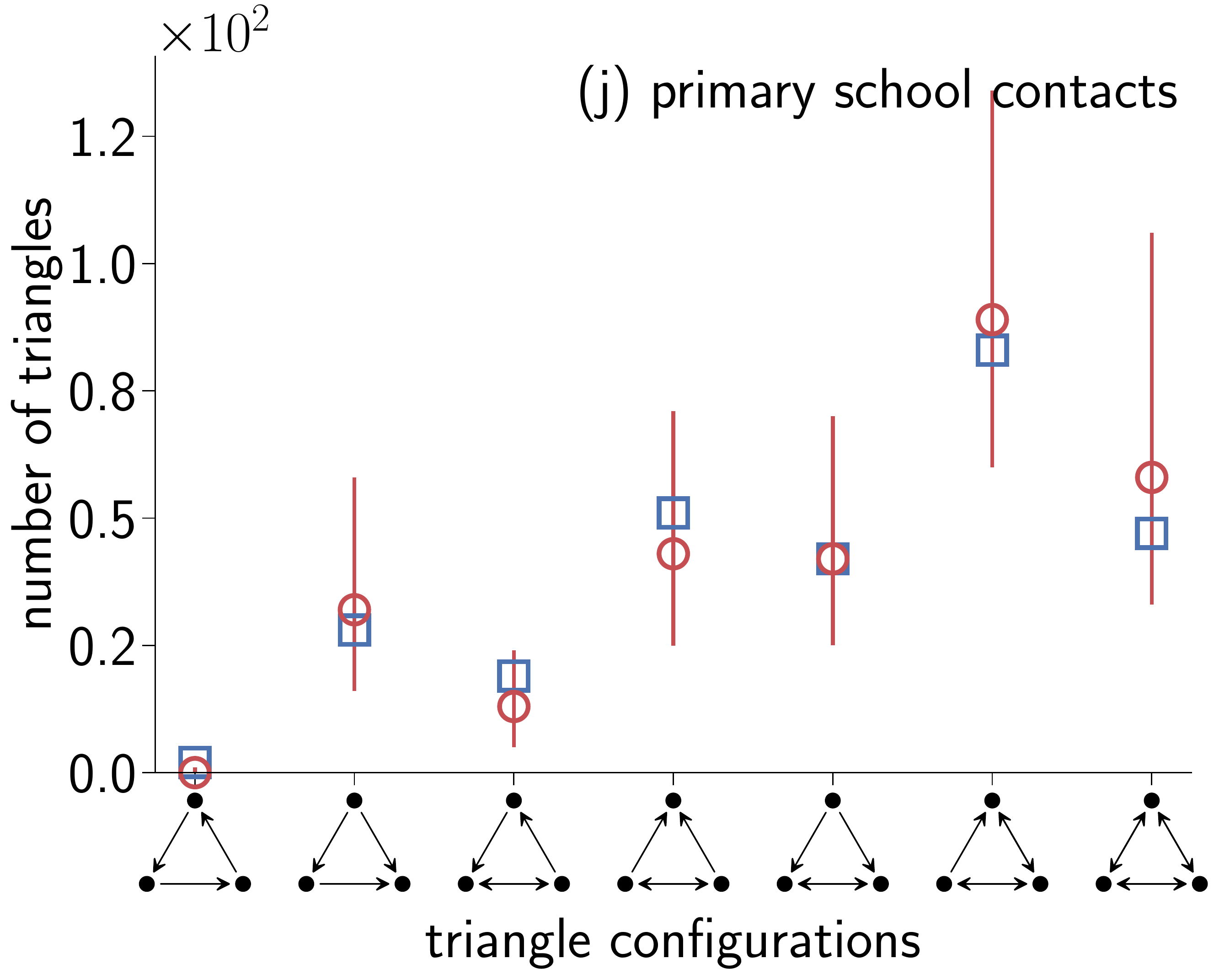}%
  \includegraphics[width=0.245\linewidth]{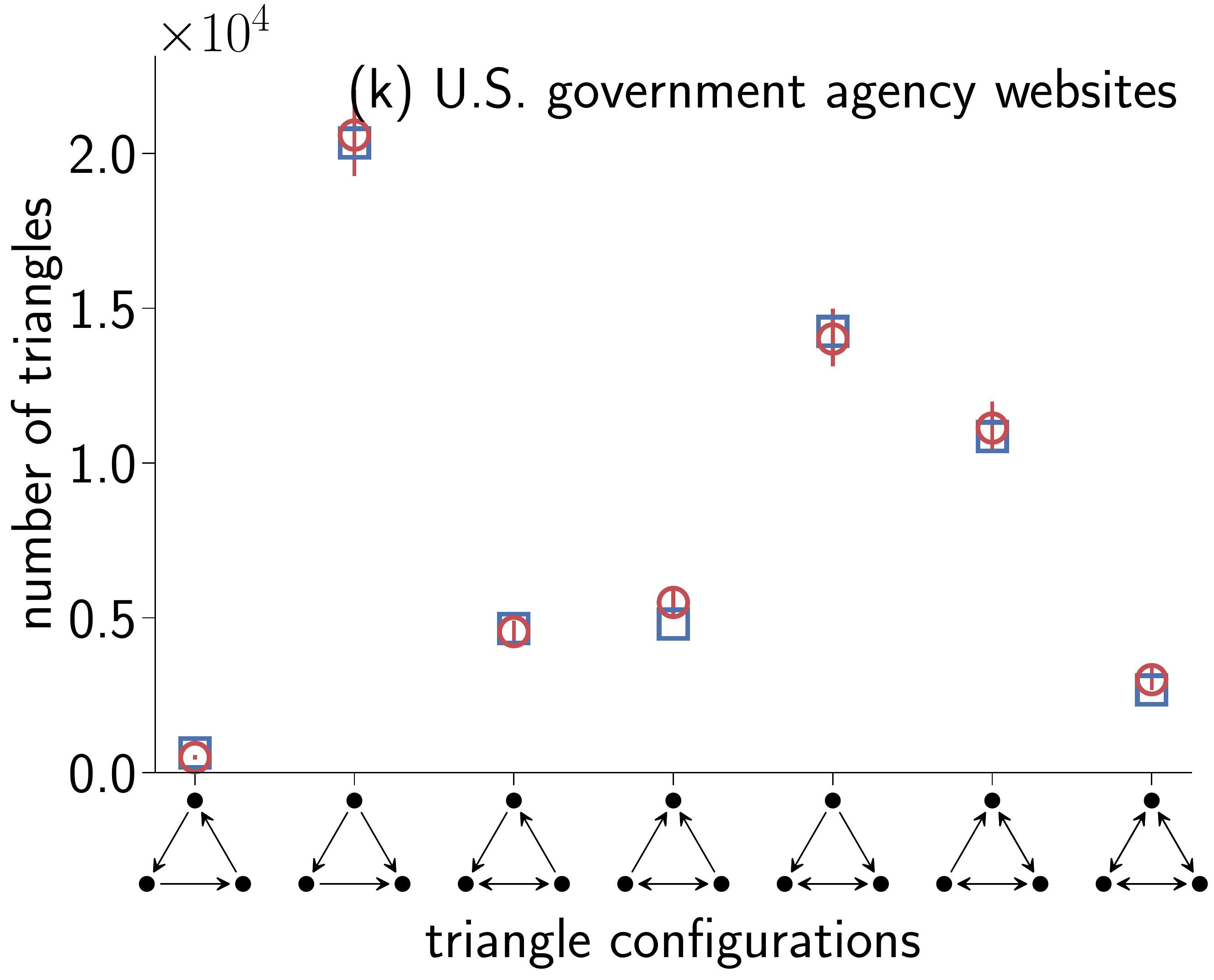}%
  \includegraphics[width=0.245\linewidth]{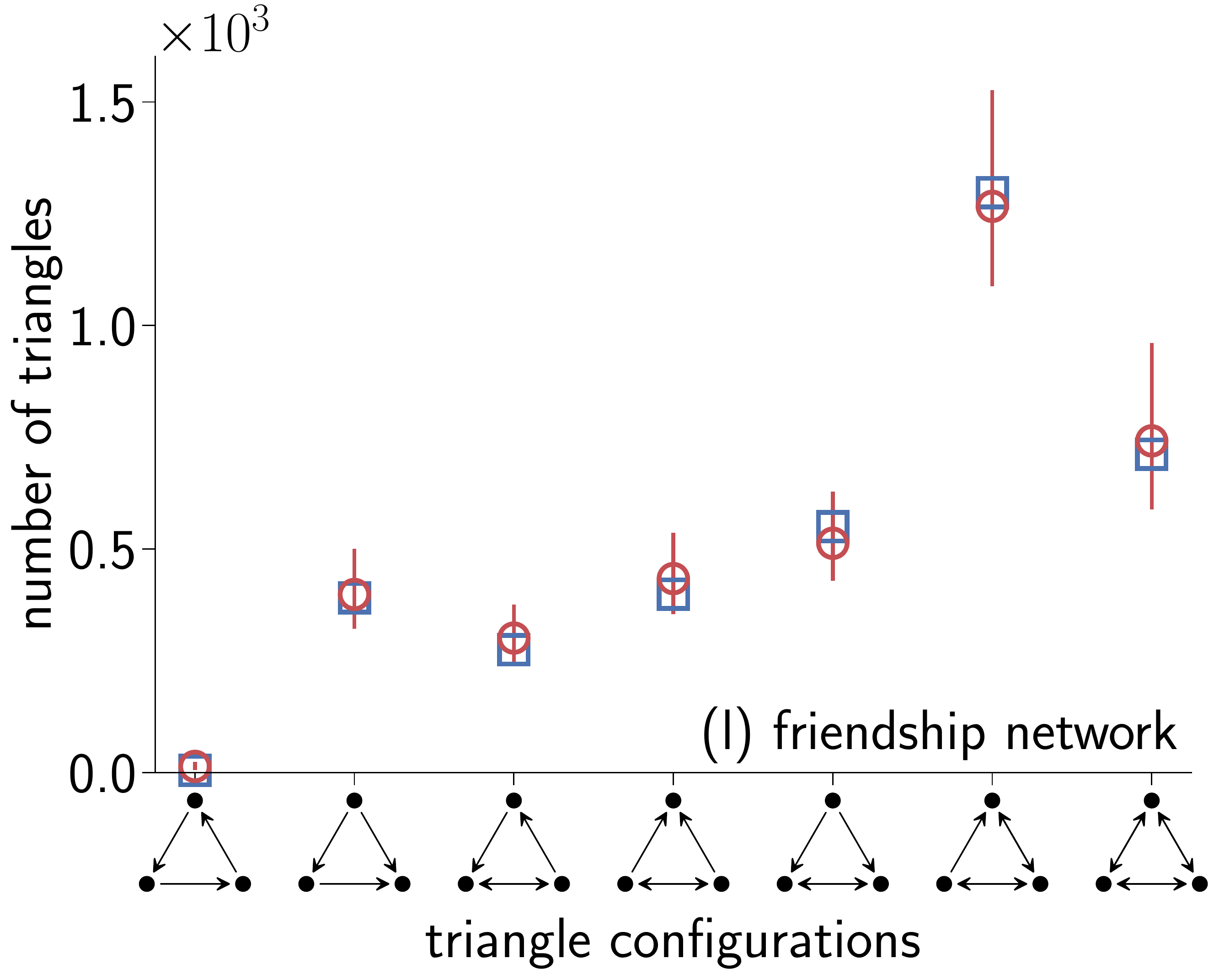}%
  \caption{%
    \textbf{Reproducing topological features of real directed networks with the directed-reciprocal $\mathbb{S}^1$ model.}
    (top row) A political blogs network (dataset:~\texttt{polblogs}~\cite{adamic2005political}).
    (a) Complementary cumulative in-degree distribution.
    (b) Complementary cumulative out-degree distribution.
    (c) In-degree and the out-degree of individual nodes.  Only a fraction of the symbols are shown to avoid cluttering the plot.
    (d) Number of triangles of each possible configuration shown in Fig.~\ref{fig:cartoon}(a).
    (middle and bottom rows) Same as (d) but for other networks.
    (e) Connectome of a tadpole larva of Ciona intestinalis (dataset:~\texttt{cintestinalis}~\cite{ryan2016cns}).
    (f) Food web of the Little Rock Lake (dataset:~\texttt{foodweb\_little\_rock}~\cite{martinez1991artifacts}).
    (g) Trade relationships between countries (dataset:~\texttt{fao\_trade}~\cite{dedomenico2015structural}).
    (h) Trust relationships among users on an online community of software developers (dataset:~\texttt{advogato}~\cite{massa2009bowling}).
    (i) Emails among employees manufacturing company (dataset:~\texttt{email\_company}~\cite{michalski2011matching}).
    (j) Friendships between high school students (dataset:~\texttt{sp\_high\_school\_diaries}~\cite{mastrandrea2015contact}).
    (k) Links between Washington State's government agencies websites (dataset:~\texttt{us\_agencies\_washington})~\cite{kosack2018functional}).
    (l) Friendships among students living in a residence hall (dataset:~\texttt{residence\_hall}~\cite{freeman1998exploring}).
    Network datasets were downloaded from The Netzschleuder network catalogue and repository (\url{https://networks.skewed.de}).
    For each dataset, the parameters of the directed-reciprocal $\mathbb{S}^1$ model were adjusted using the inference procedure described in the Supplementary Material.
    Green shaded areas in (a)--(b) and vertical lines in (c)--(l) show the estimated 95\% confidence interval (2.5 and 97.5 percentiles).
  }%
  \label{fig:real_networks}%
\end{figure*}%

\subsection{The directed-reciprocal \texorpdfstring{$\mathbb{S}^1$}{S1} model}
%
As mentioned at the beginning of the previous subsection, the directed $\mathbb{S}^1$ model generates reciprocal links by chance, that is when two directed links \textit{happen} to exist in opposite directions between a given pair of nodes.  However, although Fig.~\ref{fig:reciprocity_real} shows that reciprocity and the density of triangles are somewhat correlated in real directed complex networks, we found that relying on luck does not allow the accurate reproduction of the levels of reciprocity found in most network datasets.  In other words, once $\{\kin_i\}_{i=1,\ldots,N}$ and $\{\kout_i\}_{i=1,\ldots,N}$ have been set to reproduce the joint degree sequence and $\beta$ has been chosen to reproduce the density of triangles, an additional parameter is required to accurately tune the level of reciprocity to the one of a target real directed complex network.

The directed-reciprocal $\mathbb{S}^1$ model consists in the combination of the two aforementioned modeling approaches.  Combining Eqs.~\eqref{eq:Pij11}~and~\eqref{eq:connection_probability} fixes $P_{ij}(1,1)$, which in turn fixes $P_{ij}(1,0)$ and $P_{ij}(0,1)$ via Eqs.~(\ref{eq:marginal_conn_prob}).  Finally, asking for normalization sets $P_{ij}(0,0)$.  The parameter $\nu$ therefore corresponds to the extra parameter required to control the level of reciprocity.

Figure~\ref{fig:reciprocity_synth} illustrates the range of reciprocity that can be obtained with the directed-reciprocal $\mathbb{S}^1$ model as well as with the directed soft configuration model, which corresponds to the limit $\beta \rightarrow 0$~\cite{vanderkolk2022anomalous}.  In both panels, nodes were distributed homogeneously at random on the circle and assigned hidden degrees.  In the top panel, the in- and out-degrees are fully correlated---so that $\kout_i=\kin_i \; \forall i$---while they are uncorrelated in the bottom panel. Links were then added randomly according to the joint probabilities $P_{ij}(a_{ij}, a_{ji})$ defined by Eqs.~\eqref{eq:marginal_conn_prob},~\eqref{eq:normalization_conn_joint_prob},~\eqref{eq:Pij11},~and~\eqref{eq:connection_probability}.  Figure~\ref{fig:reciprocity_synth} illustrates the effect that both parameter $\beta$ and the correlation between $\kin$ and $\kout$ have on reciprocity (see caption for details).  Indeed, we note that stronger correlations between $\kin$ and $\kout$ and larger values of $\beta$ both yield networks with a higher reciprocity.  To understand this interplay, we introduce $\kappa_{ij} = \kout_i\kin_j$ and we use Eq.~\eqref{eq:connection_probability} to rewrite Eqs.~\eqref{eq:Pij11}--\eqref{eq:average_reciprocal_degree} as
\begin{subequations}
\begin{align}
  \label{eq:reciprocity_S1_all_nu}
  \Expected{r}
    & \approx \frac{\Expected{\drec}}{\Expected{\dout}}
      = \begin{cases}
          (1 + \nu) \ExpectedCond{r}{\nu\!=\!0} - \nu \ExpectedCond{r}{\nu\!=\!-1} \\\hfill \text{for } -1 \leq \nu \leq 0 \\ \\
          (1 - \nu) \ExpectedCond{r}{\nu\!=\!0} + \nu \ExpectedCond{r}{\nu\!=\!+1} \\\hfill \text{for }  0 \leq \nu \leq 1
        \end{cases}
\end{align}
with
\begin{align}
  \label{eq:reciprocity_S1_nu_plus1}
  \mkern-2.5mu\ExpectedCond{r}{\nu\!=\!+1}
    & \simeq\!\! \frac{1}{\Expected{\kappa}^2} \left\langle \min\Big\{\kappa_{ij},\kappa_{ji}\Big\} \right\rangle \ ,
\end{align}
\begin{align}
  \label{eq:reciprocity_S1_nu_zero}
  \ExpectedCond{r}{\nu\!=\!0}
    & \simeq  \frac{1}{\Expected{\kappa}^2} \left\langle \!\kappa_{ij} \ \kappa_{ji} \frac{\kappa_{ij}^{\beta - 1} - \kappa_{ji}^{\beta - 1}}{\kappa_{ij}^\beta - \kappa_{ji}^\beta} \right\rangle \ ,
\end{align}
and
\begin{align}
  \label{eq:reciprocity_S1_nu_minus1}
  \mkern-2.5mu\ExpectedCond{r}{\nu\!=\!-1}
    & \simeq  \!\frac{\sin(\pi/\beta)}{\Expected{\kappa}^2(\pi/\beta)} \langle f(\kappa_{ij},\kappa_{ji},\beta) \rangle \ ,
\end{align}
\end{subequations}
where $f(\kappa_{ij}, \kappa_{ji}, \beta)$ is a symmetric function with respect to its first two arguments, and an increasing function of its third.  A detailed derivation of these equations is provided in the Supplementary Material.  Equation~\eqref{eq:reciprocity_S1_all_nu} already explains the observed linear behavior with parameter $\nu$, although with two different slopes for positive or negative values.

Regarding the dependence on parameter $\beta$ and in-degree--out-degree correlations, first, we observe that Eq.~\eqref{eq:reciprocity_S1_nu_plus1} does not depend on $\beta$ and therefore that maximal reciprocity---attained at $\nu=1$---only depends on the correlation between $\kin$ and $\kout$.  This observation is confirmed in Figure~\ref{fig:reciprocity_synth}.  Equation~\eqref{eq:reciprocity_S1_nu_plus1} also confirms our previous observation that fully reciprocal networks (i.e. $r=1$) can only be expected when $P(a_{ij} = 1|\kout_i, \kin_j, \Delta\theta_{ij}) = P(a_{ji} = 1|\kout_j, \kin_i, \Delta\theta_{ij})$ which implies that $\kin$ and $\kout$ are fully correlated (i.e. $\kin_i = \kout_i$ for $i=1,\ldots, N$).  Any weaker correlation will imply a lower reciprocity since the step functions will oversample $\min\{\kappa_{ij},\kappa_{ji}\}$ leading to the right-hand side of Eq.~\eqref{eq:reciprocity_S1_nu_plus1} being lower than 1.

Second, we observe in Fig.~\ref{fig:reciprocity_synth} that larger values of $\beta$ allow for higher levels of reciprocity.  This can be understood by noting that Eq.~\eqref{eq:connection_probability} becomes a step function as $\beta \to \infty$.  In this limit, any pair of nodes $i$ and $j$ for which $\max\left\{\chi_{ij}, \chi_{ji}\right\} < 1$ will be connected by a reciprocal link with probability 1.  As $\beta$ decreases, this probability for these same pairs of nodes will also decrease, and this drop in likelihood will not be compensated by the fact that reciprocal links between pairs of nodes with larger $\chi_{ij}$ or $\chi_{ji}$ are becoming likelier [Eq.~\eqref{eq:connection_probability} decreases too quickly].  As a consequence, the reciprocity increases with $\beta$.  This relationship becomes explicit when $\kin$ and $\kout$ are fully correlated (i.e. $\kappa_{ij} = \kappa_{ji}$) as Eq.~\eqref{eq:reciprocity_S1_nu_zero} becomes $\ExpectedCond{r}{\nu\!=\!0} \simeq 1 - 1/\beta$.

\subsection{Modeling real networks}
%
We explored the capacity of the directed-reciprocal $\mathbb{S}^1$ model to reproduce the structure of real directed complex networks, most notably their level of reciprocity and their clustering patterns [see~Fig.~\ref{fig:cartoon}(a)].  Inspired by the parameter inference procedure of Ref.~\cite{garcia-perez2019mercator}, we designed an inference algorithm for the $2N+2$ parameters---$\{\kin_i,\kout_i\}_{i=1,\ldots,N}$, $\beta$ and $\nu$--- so that the directed-reciprocal $\mathbb{S}^1$ model reproduces, on average, the joint in/out-degree sequence, the reciprocity and the density of triangles (regardless of their configuration) of an original real directed complex network ($2N+2$ constraints).  These $2N+2$ parameters are inferred when averaging over all possible angular positions, meaning that angular positions $\{\theta_i\}_{i=1,\ldots,N}$ are not inferred.  A detailed description of the inference algorithm is provided in the Supplementary Material, and its implementation in C++ is publicly available (see Methods).

We ran our algorithm on more than two dozen representative datasets from The Netzschleuder network catalogue and repository (\url{https://networks.skewed.de}).  The results are shown in Fig.~\ref{fig:real_networks}.  Figures~\ref{fig:real_networks}(a)--(c) provide a representative illustration of the excellent agreement between the local properties of networks generated by our model, the in/out-degree sequence, and those of the real counterpart.  Beyond the degree sequences, Fig.~\ref{fig:real_networks}(c) shows that the model reproduces the observed correlations between in- and out-degrees.  The most striking result, however, consists in the accuracy with which the directed-reciprocal $\mathbb{S}^1$ model can reproduce the variety of clustering patterns observed in a wide range of real directed complex networks.  Indeed, Figs.~\ref{fig:real_networks}(d)--(l)---as well as in the Supplementary Material---show that only two parameters are enough to match the observed reciprocity and nontrivial clustering patterns, thereby implying that clustering in directed networks arises as a consequence of geometry and of the tendency to generate reciprocated interactions.

\section{Discussion}
%
Asymmetric interactions within complex systems are the norm rather than the exception~\cite{asllani2018structure}.  Yet, and for lack of sufficiently adequate modeling frameworks, it is common to see directionality neglected and somewhat treated as an ``afterthought''~\cite{johnson2020digraphs}; the underlying assumption being that the undirected representation of many complex systems encodes most of the relationship between the behavior of these systems and their structure.  Mounting evidence argues that this is not the case, however, and that directionality drastically impacts the global organization and the behavior of these systems~\cite{asllani2018structure, coletta2020network, duan2022network, johnson2014trophic, johnson2017looplessness, johnson2020digraphs, klaise2016neurons, nicolaou2020nonnormality, qu2014nonconsensus, shao2009dynamic}.  Overlooking directionality therefore provides an incomplete picture when not a misleading one.

Extending the framework of network geometry to directed networks has therefore been an urgent matter for many years, but progress was impeded by the fundamental incompatibility between asymmetric interactions and the symmetry of distances in any metric space.  In this paper, we have shown that this incompatibility can be bypassed by rethinking the relationship between connections and distances.  Doing so results in a powerful and versatile framework, amenable to several analytical calculations, and that can be easily adjusted to reproduce several properties observed in a large variety of network datasets.

Most importantly, we have shown that our framework reproduces the intricate patterns of reciprocity and clustering observed in real complex directed networks.  Albeit local, these features have a significant impact on the global behavior of these networks.  For instance, they affect the outcome of spreading dynamics~\cite{klaise2016neurons}, impact the stability of food webs~\cite{johnson2014trophic, johnson2017looplessness}, and play a central role for flexible navigation and context-dependent action selection in connectomes~\cite{hulse2021connectome}.  Also, the information encoded in the patterns of reciprocity and of clustering is rich enough for them to act as a signature of the nature of real complex networks (social, technological, physical, biological, etc.)~\cite{ahnert2008clustering, garlaschelli2004patterns, jia2021directed}.  It is therefore paramount for any realistic modeling approach to be able to reproduce these intricate patterns of reciprocity and clustering.  Now that the gap between asymmetric interactions and symmetric metric distances has been bridged, accurate modeling of a wide and diverse range of complex systems is now within reach.

%


\clearpage
\small
\vspace{1.5\baselineskip}\par%
\noindent\textbf{\normalsize Methods}\\
%
%
\noindent\textbf{Density of triangles in directed networks.}
We quantify the density of triangles in a directed network with the average local clustering coefficient, $\bar{c}_\mathrm{undir}$, computed using the undirected version of the original directed network.  From the adjacency matrix of the directed network $\mathbf{A}=\{a_{ij}\}$, we define the undirected adjacency matrix $\mathbf{\tilde{A}}$ whose elements are $\tilde{a}_{ij} = \max(a_{ij},a_{ji})$.  The density of triangles is then
\begin{align}
  \bar{c}_\mathrm{undir}
    = \frac{1}{N} \sum_{i=1}^{N} \frac{2T_i}{k_i (k_i - 1)} \mathbbm{1}_{\{k_i>1\}}
\end{align}
where $T_i = \frac{1}{2}[\mathbf{\tilde{A}}^3]_{ii}$ is the number of triangles to which node $i$ participates, $k_i = \sum_{j=1}^{N} [\mathbf{\tilde{A}}]_{ij}$ is the degree of node $i$, and $\mathbbm{1}_{\{\cdot\}}$ is the indicator function.
\vspace{0.5\baselineskip}\par%
\noindent\textbf{Directed soft configuration model.}
The directed soft configuration model is the unique ensemble of unbiased sparse random graphs whose entropy is maximized across all graphs with a given expected joint in- and out-degree distribution~\cite{bianconi2009entropy, vanderhoorn2018sparse}.  It consists in $N$ nodes, each of which is assigned a pair of hidden degrees $\kin$ and $\kout$ according to $\rho(\kin, \kout)$. In this model, a directed link from node $i$ to node $j$ exists with probability
\begin{align} \label{eq:directed_cm_connection_probability}
  P(a_{ij} = 1 | \kout_i, \kin_j) = \frac{1}{1 + \frac{N \Expected{\kappa}}{\kout_i \kin_j}} \simeq \frac{\kout_i \kin_j}{N \Expected{\kappa}}  \ ,
\end{align}
where the approximation holds in the sparse limit. Note that the directed $\mathbb{S}^1$ model falls back on the directed soft configuration model in the limit $\beta \to 0$~\cite{boguna2020small}.
\vspace{0.5\baselineskip}\par%
\noindent\textbf{Correspondence with the directed soft configuration model.}
To see how the directed $S^1$ model falls back on the directed soft configuration model, we first average Eq.~\eqref{eq:connection_probability} over the angular distance $\Delta\theta_{ij}$ to obtain the expected probability for a link to exist from node $i$ to node $j$ in the network ensemble
\begin{align} \label{eq:expected_connection_probability}
  \ExpectedCond{a_{ij}}{\kout_i, \kin_j}
    ={}_2F_1\left(1,\frac{1}{\beta},1+\frac{1}{\beta},-\left( \frac{N}{2 \mu \kout_i \kin_j }\right)^\beta \right)\ ,
\end{align}
where ${}_2F_1$ is the hypergeometric function. From this expression, we show in the Supplementary Material that in the limit $N/(\kout_i\kin_j) \to \infty$ the average connection probability becomes
\begin{align} \label{eq:expected_connection_probability2}
  \ExpectedCond{a_{ij}}{\kout_i, \kin_j} \simeq \frac{\kout_i \kin_j}{N \Expected{\kappa}} \ ,
\end{align}
which we identify as the connection probability of the sparse directed soft configuration model, Eq.~\eqref{eq:directed_cm_connection_probability}.
\vspace{0.5\baselineskip}\par%
\noindent\textbf{Directed-reciprocal soft configuration model.}
Akin to the directed-reciprocal $\mathbb{S}^1$ model, the directed-reciprocal soft configuration model is a combination of the framework controlling reciprocity of Sec.~\ref{sec:reciprocity} and of the directed-reciprocal soft configuration model presented above (which provides the marginal probabilities).
\vspace{0.5\baselineskip}\par%
\noindent\textbf{Network datasets.}
The list of all datasets is provided in the Supplementary Material.

\vspace{1.5\baselineskip}\par%
\noindent\textbf{\normalsize Data availability}\\
The network datasets used in the article have been made publicly available by the original authors and were downloaded from The Netzschleuder network catalogue and repository (\url{https://networks.skewed.de}).\\

\vspace{1.0\baselineskip}\par%
\noindent\textbf{\normalsize Code availability}\\
The scripts and the source code of the programs used to produce the figures will be available at \url{https://github.com/networkgeometry/directed-geometric-networks}.

\vspace{1.5\baselineskip}\par%
\noindent\textbf{\normalsize Acknowledgements}\\
The authors are grateful to Louis J. Dub\'e for comments.  A.A. acknowledges financial support from the Sentinelle Nord initiative of the Canada First Research Excellence Fund and from the Natural Sciences and Engineering Research Council of Canada (project 2019-05183). M.~A.~S. and M.~B. acknowledge support from: Grant TED2021-129791B-I00 funded by MCIN/AEI/10.13039/501100011033 and the ``European Union NextGenerationEU/PRTR''; Grant PID2019-106290GB-C22 funded by MCIN/AEI/10.13039/501100011033; and Generalitat de Catalunya grant number 2021SGR00856. M. B. acknowledges the ICREA Academia award, funded by the Generalitat de Catalunya.

\ifarXiv
    

\section*{Supplementary material}

# Geometric description of clustering in directed networks
## —Supplementary material—

Antoine Allard,[1,2] M. Ángeles Serrano,[3,4,5] and Marián Boguñá[3,4]

[1] *Département de physique, de génie physique et d'optique,*
*Université Laval, Québec (Québec), Canada G1V 0A6*

[2] *Centre interdisciplinaire en modélisation mathématique,*
*Université Laval, Québec (Québec), Canada G1V 0A6*

[3] *Departament de Física de la Matèria Condensada,*
*Universitat de Barcelona, Martí i Franquès 1, E-08028 Barcelona, Spain*

[4] *Universitat de Barcelona Institute of Complex Systems (UBICS), Universitat de Barcelona, Barcelona, Spain*

[5] *Institució Catalana de Recerca i Estudis Avançats (ICREA),*
*Passeig Lluís Companys 23, E-08010 Barcelona, Spain*

(Dated: February 17, 2023)

This document provides further and complementary details about the results presented and discussed in the main text.

**CONTENTS**

## S.I. CONTROLLING THE RECIPROCITY IN RANDOM DIRECTED NETWORKS

We consider a general random directed networks model in which $p_{ij}$ is the probability for a directed link to exist from node $i$ to node $j$. We denote the number of nodes with $N$. To control the level of reciprocity, our approach focuses on each *pair* of directed links between two nodes and defines the four following symmetrical probabilities

$$P_{ij}(a_{ij} = 0, a_{ji} = 0) \quad \text{(none of the two possible directed links exist)}$$
$$P_{ij}(a_{ij} = 1, a_{ji} = 0) \quad \text{(the link from node } i \text{ to node } j \text{ exists but the other does not)}$$
$$P_{ij}(a_{ij} = 0, a_{ji} = 1) \quad \text{(the link from node } j \text{ to node } i \text{ exists but the other does not)}$$
$$P_{ij}(a_{ij} = 1, a_{ji} = 1) \quad \text{(both directed links exists)}$$

with $1 \leq i < j \leq N$ and where $a_{ij}$ is 1 if there is a directed link from node $i$ to node $j$, and 0 otherwise. These four joint probabilities are normalized

$$P_{ij}(a_{ij} = 0, a_{ji} = 0) + P_{ij}(a_{ij} = 1, a_{ji} = 0) + P_{ij}(a_{ij} = 0, a_{ji} = 1) + P_{ij}(a_{ij} = 1, a_{ji} = 1) = 1 , \tag{S1}$$

and their marginals must be coherent with the random directed network model

$$P_{ij}(a_{ij} = 1, a_{ji} = 0) + P_{ij}(a_{ij} = 1, a_{ji} = 1) = p_{ij} , \tag{S2a}$$
$$P_{ij}(a_{ij} = 0, a_{ji} = 1) + P_{ij}(a_{ij} = 1, a_{ji} = 1) = p_{ij} . \tag{S2b}$$

To connect the probabilities $P_{ij}(a_{ij}, a_{ji})$ with the reciprocity in the network ensemble defined by the model, we look at the following correlation coefficient

$$\rho_{ij} = \frac{\langle a_{ij} a_{ji} \rangle - \langle a_{ij} \rangle \langle a_{ji} \rangle}{\sqrt{(\langle a_{ij}^2 \rangle - \langle a_{ij} \rangle^2)(\langle a_{ji}^2 \rangle - \langle a_{ji} \rangle^2)}} = \frac{P_{ij}(1,1) - p_{ij} p_{ji}}{\sqrt{p_{ij}(1 - p_{ij}) p_{ji}(1 - p_{ji})}} . \tag{S3}$$

for each pair $(i, j)$ with $1 \leq i < j \leq N$, and where $\langle \cdot \rangle$ corresponds to an average over the network ensemble. A closed form for $P_{ij}(1, 1)$ in terms of $p_{ij}$ and $p_{ji}$ can be obtained by combining Eqs. (S2)–(S3) alongside the requirement that each of the four joint probabilities $P_{ij}(a_{ij}, a_{ji})$ is bounded in $[0, 1]$:

1. From Eq. (S3), we can isolate

$$P_{ij}(a_{ij} = 1, a_{ji} = 1) = p_{ij} p_{ji} + \rho_{ij} \sqrt{p_{ij}(1 - p_{ij}) p_{ji}(1 - p_{ji})} , \tag{S4}$$

which will be bounded in $[0, 1]$ if

$$-\frac{p_{ij} p_{ji}}{\sqrt{p_{ij}(1 - p_{ij}) p_{ji}(1 - p_{ji})}} \leq \rho_{ij} \leq \frac{1 - p_{ij} p_{ji}}{\sqrt{p_{ij}(1 - p_{ij}) p_{ji}(1 - p_{ji})}} . \tag{S5}$$

2. Combining Eqs. (S2a) and (S4), we can isolate

$$P_{ij}(a_{ij} = 1, a_{ji} = 0) = p_{ij} - P(a_{ij} = 1, a_{ji} = 1) = p_{ij}(1 - p_{ji}) - \rho_{ij} \sqrt{p_{ij}(1 - p_{ij}) p_{ji}(1 - p_{ji})} \tag{S6}$$

which will be bounded in $[0, 1]$ if

$$\frac{p_{ij}(1 - p_{ji}) - 1}{\sqrt{p_{ij}(1 - p_{ij}) p_{ji}(1 - p_{ji})}} \leq \rho_{ij} \leq \frac{p_{ij}(1 - p_{ji})}{\sqrt{p_{ij}(1 - p_{ij}) p_{ji}(1 - p_{ji})}} . \tag{S7}$$

3. Combining Eqs. (S2b) and (S4), we can isolate

$$P_{ij}(a_{ij} = 0, a_{ji} = 1) = p_{ji} - P_{ij}(a_{ij} = 1, a_{ji} = 1) = p_{ji}(1 - p_{ij}) - \rho_{ij} \sqrt{p_{ij}(1 - p_{ij}) p_{ji}(1 - p_{ji})} \tag{S8}$$

which will be bounded in $[0, 1]$ if

$$\frac{p_{ji}(1 - p_{ij}) - 1}{\sqrt{p_{ij}(1 - p_{ij}) p_{ji}(1 - p_{ji})}} \leq \rho_{ij} \leq \frac{p_{ji}(1 - p_{ij})}{\sqrt{p_{ij}(1 - p_{ij}) p_{ji}(1 - p_{ji})}} . \tag{S9}$$

4. Combining Eqs. (S1), (S4), (S6) and (S8), we can isolate

$$P_{ij}(a_{ij}=0, a_{ji}=0) = 1 - P_{ij}(a_{ij}=1, a_{ji}=0) - P_{ij}(a_{ij}=0, a_{ji}=1) - P_{ij}(a_{ij}=1, a_{ji}=1)$$
$$= (1-p_{ji})(1-p_{ij}) + \rho_{ij}\sqrt{p_{ij}(1-p_{ij})p_{ji}(1-p_{ji})} \qquad (S10)$$

which will be bounded in $[0,1]$ if

$$-\frac{(1-p_{ji})(1-p_{ij})}{\sqrt{p_{ij}(1-p_{ij})p_{ji}(1-p_{ji})}} \leq \rho_{ij} \leq \frac{1-(1-p_{ji})(1-p_{ij})}{\sqrt{p_{ij}(1-p_{ij})p_{ji}(1-p_{ji})}} \ . \qquad (S11)$$

The lower and upper bounds for $\rho_{ij}$ are therefore

$$\rho_{ij}^{\min} = \frac{1}{\sqrt{p_{ij}(1-p_{ij})p_{ji}(1-p_{ji})}}\max\left\{-p_{ij}p_{ji},\ p_{ij}(1-p_{ji})-1,\ p_{ji}(1-p_{ij})-1,\ -(1-p_{ji})(1-p_{ij})\right\}$$
$$= \begin{cases} -\dfrac{p_{ij}p_{ji}}{\sqrt{p_{ij}(1-p_{ij})p_{ji}(1-p_{ji})}} & \text{if } p_{ij}+p_{ji} < 1 \\ -\dfrac{(1-p_{ji})(1-p_{ij})}{\sqrt{p_{ij}(1-p_{ij})p_{ji}(1-p_{ji})}} & \text{if } p_{ij}+p_{ji} > 1 \end{cases} \qquad (S12)$$

and

$$\rho_{ij}^{\max} = \frac{1}{\sqrt{p_{ij}(1-p_{ij})p_{ji}(1-p_{ji})}}\min\left\{1-p_{ij}p_{ji},\ p_{ij}(1-p_{ji}),\ p_{ji}(1-p_{ij}),\ 1-(1-p_{ji})(1-p_{ij})\right\}$$
$$= \begin{cases} \dfrac{p_{ij}(1-p_{ji})}{\sqrt{p_{ij}(1-p_{ij})p_{ji}(1-p_{ji})}} & \text{if } p_{ij} < p_{ji} \\ \dfrac{p_{ji}(1-p_{ij})}{\sqrt{p_{ij}(1-p_{ij})p_{ji}(1-p_{ji})}} & \text{if } p_{ij} > p_{ji} \ . \end{cases} \qquad (S13)$$

To control the level of reciprocity, we introduce a parameter $\nu \in [-1,1]$ controlling $\rho_{ij}$ such that $\rho_{ij}^{\min} \leq \rho_{ij} \leq \rho_{ij}^{\max}$

$$\rho_{ij} = \begin{cases} |\nu|\rho_{ij}^{\min} & \text{if } -1 \leq \nu \leq 0 \\ |\nu|\rho_{ij}^{\max} & \text{if } \ \ \ 0 \leq \nu \leq 1 \end{cases} \ . \qquad (S14)$$

Substituting Eq. (S14) into Eq. (S4) allows us to isolate

$$P_{ij}(a_{ij}=1, a_{ji}=1) = \begin{cases} (1+\nu)p_{ij}p_{ji} - \nu(p_{ij}+p_{ji}-1)H(p_{ij}+p_{ji}-1) & -1 \leq \nu \leq 0 \\ (1-\nu)p_{ij}p_{ji} + \nu \min\left\{p_{ij}, p_{ji}\right\} & 0 \leq \nu \leq 1 \end{cases}, \qquad (S15)$$

where $H(\cdot)$ is the Heaviside step function. This last equation alongside Eqs. (S1) and (S2) complete the approach for controlling reciprocity, whose level is tuned by the parameter $\nu$.

## S.II. ANALYSIS OF THE DIRECTED $\mathbb{S}^1$ MODEL

### A. Description of the model

We consider $N$ nodes positioned on a circle of radius $R = N/2\pi$, thus setting the density of nodes to 1 without loss of generality. Each node $i$ is independently and identically assigned an angular position $\theta_i$ and a pair of *hidden degrees*, $\kappa_i^-$ and $\kappa_i^+$ which, as shown below, are related to their in- and out-degree, respectively. The angular positions are scattered on the circle according to the uniform probability density function (pdf)

$$\varphi(\theta) = \frac{1}{2\pi} . \tag{S16}$$

The hidden degrees are also assigned randomly according to the joint pdf $\rho(\kappa^-, \kappa^+)$, whose only constraint is

$$\iint \kappa^- \rho(\kappa^-, \kappa^+) d\kappa^- d\kappa^+ = \langle \kappa^- \rangle \equiv \langle \kappa \rangle \tag{S17a}$$

$$\iint \kappa^+ \rho(\kappa^-, \kappa^+) d\kappa^- d\kappa^+ = \langle \kappa^+ \rangle \equiv \langle \kappa \rangle . \tag{S17b}$$

A directed link exists from node $i$ to node $j$ with probability

$$P(a_{ij} = 1 | \kappa_i^+, \kappa_j^-, \Delta\theta_{ij}) = \frac{1}{1 + \chi_{ij}^\beta} \quad \text{with} \quad \chi_{ij} = \frac{R \Delta\theta_{ij}}{\mu \kappa_i^+ \kappa_j^-} = \frac{N \Delta\theta_{ij}}{2\pi \mu \kappa_i^+ \kappa_j^-} \tag{S18}$$

where $\Delta\theta_{ij} = \Delta\theta_{ji} = \pi - |\pi - |\theta_i - \theta_j||$ is the minimal angular distance between nodes $i$ and $j$, and where $\mu > 0$ and $\beta > 1$ are parameters of the model. Note that we will omit writing explicitly the dependency over $\beta$, $\mu$ and $N$ for brevity. Note also that $\varphi(\theta) = \frac{1}{2\pi}$ implies that the pdf for $\Delta\theta_{ij}$ is simply $1/\pi$.

### B. Out-degree of nodes

Let us first consider $N$ nodes, each of which has been assigned an angular position $\theta$, a hidden in-degree $\kappa^-$ and a hidden out-degree $\kappa^+$. The sequence of angular positions, noted $\boldsymbol{\theta} \equiv \{\theta_1, \ldots, \theta_N\}$ is distributed according to the pdf $\prod_{i=1}^{N} \varphi(\theta_i) = (2\pi)^{-N}$, and the hidden degrees sequence, noted $\boldsymbol{\kappa} \equiv \{\kappa_1^-, \kappa_1^+, \ldots, \kappa_N^-, \kappa_N^+\}$ is distributed according to the pdf $\prod_{i=1}^{N} \rho(\kappa_i^-, \kappa_i^+)$. The connection probability given by Eq. (S18) alongside the sequences $\boldsymbol{\theta}$ and $\boldsymbol{\kappa}$ define a random network ensemble in which node $i$ has a out-degree equal to $k_i^+$ with probability $P_i^+(k_i^+|\boldsymbol{\theta}, \boldsymbol{\kappa})$. The associated probability generating function (pgf) is defined as

$$H_i^+(z|\boldsymbol{\kappa}, \boldsymbol{\theta}) = \sum_{k_i^+=0}^{N-1} P_i^+(k_i^+|\boldsymbol{\theta}, \boldsymbol{\kappa}) z^{k_i^+} = \prod_{\substack{j=1 \\ j \neq i}}^{N} \left[ 1 - P(a_{ij} = 1|\kappa_i^+, \kappa_j^-, \Delta\theta_{ij}) + z P(a_{ij} = 1|\kappa_i^+, \kappa_j^-, \Delta\theta_{ij}) \right], \tag{S19}$$

where we used the fact that the existence of each link is *conditionally* independent from the existence of the others (i.e. they are independent events given the hidden variables $\theta_i$ and $\kappa_i^+$). General expressions for the expected out-degree of node $i$, its variance and for the ensemble average out-degree are respectively

$$\langle k_i^+ | \boldsymbol{\kappa}, \boldsymbol{\theta} \rangle = \left. \frac{\partial H_i^+(z|\boldsymbol{\kappa}, \boldsymbol{\theta})}{\partial z} \right|_{z=1} = \sum_{\substack{j=1 \\ j \neq i}}^{N} P(a_{ij} = 1 | \kappa_i^+, \kappa_j^-, \Delta\theta_{ij}) , \tag{S20}$$

$$\text{Var}\left[k_i^+ | \boldsymbol{\kappa}, \boldsymbol{\theta}\right] = \left. \frac{\partial^2 H_i^+(z|\boldsymbol{\kappa}, \boldsymbol{\theta})}{\partial z^2} \right|_{z=1} + \left. \frac{\partial H_i^+(z|\boldsymbol{\kappa}, \boldsymbol{\theta})}{\partial z} \right|_{z=1} - \left[ \left. \frac{\partial H_i^+(z|\boldsymbol{\kappa}, \boldsymbol{\theta})}{\partial z} \right|_{z=1} \right]^2$$

$$= \sum_{\substack{j=1 \\ j \neq i}}^{N} P(a_{ij} = 1|\kappa_i^+, \kappa_j^-, \Delta\theta_{ij})[1 - P(a_{ij} = 1|\kappa_i^+, \kappa_j^-, \Delta\theta_{ij})] , \tag{S21}$$

and

$$\langle k^+|\boldsymbol{\kappa},\boldsymbol{\theta}\rangle = \frac{1}{N}\sum_{i=1}^{N}\langle k_i^+|\boldsymbol{\kappa},\boldsymbol{\theta}\rangle = \frac{1}{N}\sum_{i=1}^{N}\sum_{\substack{j=1\\j\neq i}}^{N}P(a_{ij}=1|\kappa_i^+,\kappa_j^-,\Delta\theta_{ij})\,. \tag{S22}$$

Let us now zoom out of the random network ensemble defined by specific sequences $\boldsymbol{\theta}$ and $\boldsymbol{\kappa}$ to focus instead on the random network ensemble defined by the pdfs $\varphi(\theta)$ and $\rho(\kappa^-,\kappa^+)$ (i.e. any sequences $\boldsymbol{\theta}$ of length $N$ and $\boldsymbol{\kappa}$ of length $2N$ drawn from their respective pdf). Averaging over all angular positions (or, equivalently, over all angular distances), the expected probability for a link to exist from node $i$ to node $j$ in the network ensemble becomes

$$\begin{aligned}
\langle a_{ij}|\kappa_i^+,\kappa_j^-\rangle &= \frac{1}{\pi}\int_0^\pi P(a_{ij}=1|\kappa_i^+,\kappa_j^-,\Delta\theta_{ij})d\Delta\theta_{ij}\\
&= \frac{1}{\pi}\int_0^\pi \frac{1}{1+\chi_{ij}^\beta}d\Delta\theta_{ij}\\
&= \frac{2\mu\kappa_i^+\kappa_j^-}{N}\int_0^{\frac{N}{2\mu\kappa_i^+\kappa_j^-}}\frac{1}{1+\chi_{ij}^\beta}d\chi_{ij}\\
&= {}_2F_1\left(1,\frac{1}{\beta};1+\frac{1}{\beta};-\left[\frac{N}{2\mu\kappa_i^+\kappa_j^-}\right]^\beta\right) \tag{S23a}\\
&\simeq \frac{2\pi\mu\kappa_i^+\kappa_j^-}{\beta N\sin(\pi/\beta)}\\
&= \frac{\kappa_i^+\kappa_j^-}{N\langle\kappa\rangle}\,, \tag{S23b}
\end{aligned}$$

where $\simeq$ denotes an approximation that becomes exact in the limit $N/(\kappa_i^+\kappa_j^-)\to\infty$ [see Eqs. (S95) and (S104)], and where we set $\mu = \frac{\beta\sin(\pi/\beta)}{2\pi\langle\kappa\rangle}$ in the last equality. Averaging Eq. (S19) over every possible sequence $\boldsymbol{\theta}$ and $\boldsymbol{\kappa}\setminus\{\kappa_i^+\}$ then yields [note that $H_i^+(z|\boldsymbol{\kappa},\boldsymbol{\theta})$ does not depend on $\kappa_i^-$]

$$\begin{aligned}
H_i^+(z|\kappa_i^+) &= \int\cdots\int H_i^+(z|\boldsymbol{\kappa},\boldsymbol{\theta})\prod_{\substack{j=1\\j\neq i}}^{N}\frac{1}{\pi}d\Delta\theta_{ij}\rho(\kappa_j^-,\kappa_j^+)d\kappa_j^- d\kappa_j^+\\
&= \prod_{\substack{j=1\\j\neq i}}^{N}\Big[\iiint\Big[1-P(a_{ij}=1|\kappa_i^+,\kappa_j^-,\Delta\theta_{ij})\\
&\qquad\qquad + zP(a_{ij}=1|\kappa_i^+,\kappa_j^-,\Delta\theta_{ij})\Big]\frac{1}{\pi}d\Delta\theta_{ij}\rho(\kappa_j^-,\kappa_j^+)d\kappa_j^- d\kappa_j^+\Big]\\
&= \prod_{\substack{j=1\\j\neq i}}^{N}\Big[1-\langle a_{i\bullet}|\kappa_i^+\rangle + z\langle a_{i\bullet}|\kappa_i^+\rangle\Big]\\
&= \Big[1-\langle a_{i\bullet}|\kappa_i^+\rangle + z\langle a_{i\bullet}|\kappa_i^+\rangle\Big]^{N-1}\,, \tag{S24}
\end{aligned}$$

where

$$\langle a_{i\bullet}|\kappa_i^+\rangle = \iint\langle a_{ij}|\kappa_i^+,\kappa_j^-\rangle\rho(\kappa_j^-,\kappa_j^+)d\kappa_j^- d\kappa_j^+ \simeq \frac{2\pi\mu\langle\kappa\rangle\kappa_i^+}{\beta N\sin(\pi/\beta)} = \frac{\kappa_i^+}{N} \tag{S25}$$

is the average probability for the existence of any outgoing link from node $i$. From Eq. (S24), we conclude that the out-degree of node $i$ in the random networks ensemble will be distributed according to a binomial distribution with average

$$\langle k_i^+|\kappa_i^+\rangle = (N-1)\langle a_{i\bullet}|\kappa_i^+\rangle \simeq \frac{2\pi\mu\langle\kappa\rangle\kappa_i^+}{\beta\sin(\pi/\beta)} = \kappa_i^+\,, \tag{S26}$$

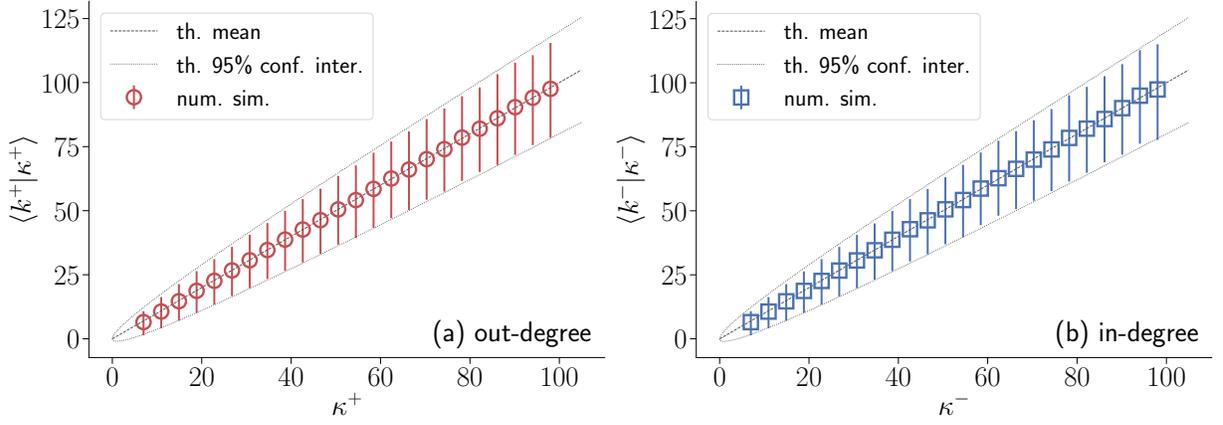

FIG. S1. **Validation of Eqs.** (S26), (S27), (S32) **and** (S33) **using numerical simulations.** Both $\kappa^-$ and $\kappa^+$ were independently and identically drawn from the pdf $\rho(\kappa) \propto \kappa^{-2.5}$ with $5 < \kappa < 100$. Symbols show $\langle k^-|\kappa^-\rangle$ and $\langle k^+|\kappa^+\rangle$ estimated from 100 random synthetic networks with $N = 25000$. Only a fraction of the symbols are shown to avoid cluttering the plot. Error bars show the estimated 95% confidence interval.

and variance

$$\text{Var}\left[k_i^+|\kappa_i^+\right] = (N-1)\langle a_{i\bullet}|\kappa_i^+\rangle\left(1 - \langle a_{i\bullet}|\kappa_i^+\rangle\right) \simeq \frac{2\pi\mu\langle\kappa\rangle\kappa_i^+}{\beta\sin(\pi/\beta)} = \kappa_i^+ \ . \tag{S27}$$

Finally, the average out-degree in the ensemble of random networks is

$$\langle k^+\rangle = \iint \langle k_i^+|\kappa_i^+\rangle \rho(\kappa_i^-,\kappa_i^+)d\kappa_i^- d\kappa_i^+ \simeq \frac{2\pi\mu\langle\kappa\rangle^2}{\beta\sin(\pi/\beta)} = \langle\kappa\rangle \ . \tag{S28}$$

As $N/(\kappa^+\kappa^-) \to \infty$, the relative fluctuations around the expected out-degree, $\sqrt{\text{Var}\left[k_i^+|\kappa_i^+\right]}/\langle k_i^+|\kappa_i^+\rangle$, will fall as $1/\sqrt{\kappa_i^+}$ and will become negligible for high out-degree nodes. The binomial distribution obtained in Eq. (S24) can therefore be approximated by a Poisson distribution in this limit

$$H_i^+(z|\kappa_i^+) = \sum_{k_i^+=0}^{N-1} P_i^+(k_i^+|\kappa_i^+)z^{k_i^+} \simeq \left[1 + (z-1)\langle a_{i\bullet}|\kappa_i^+\rangle\right]^{N-1} = \left[1 + (z-1)\frac{\kappa_i^+}{N}\right]^{N-1} \simeq \sum_{k_i^+=0}^{\infty} \frac{[\kappa_i^+]^{k_i^+}\,\mathrm{e}^{-\kappa_i^+}}{k_i^+!}z^{k_i^+} \ , \tag{S29}$$

where we identify

$$P_i^+(k_i^+|\kappa_i^+) \simeq \frac{[\kappa_i^+]^{k_i^+}\,\mathrm{e}^{-\kappa_i^+}}{k_i^+!} \tag{S30}$$

as the probability for node $i$ with hidden out-degree $\kappa_i^+$ to have a degree equal to $k_i^+$.

### C. In-degree of nodes

Repeating the same steps from the previous section yields an expression similar to Eq. (S25) for the average probability for the existence of any incoming link into node $i$

$$\langle a_{\bullet i}|\kappa_i^-\rangle = \iint \langle a_{ji}|\kappa_j^+,\kappa_i^-\rangle \rho(\kappa_j^-,\kappa_j^+)d\kappa_j^- d\kappa_j^+ \simeq \frac{2\pi\mu\langle\kappa\rangle\kappa_i^-}{\beta N\sin(\pi/\beta)} = \frac{\kappa_i^-}{N} \tag{S31}$$

an expression similar to Eq. (S26) for the expected in-degree of nodes

$$\langle k_i^-|\kappa_i^-\rangle \simeq \frac{2\pi\mu\langle\kappa\rangle\kappa_i^-}{\beta\sin(\pi/\beta)} = \kappa_i^- \ , \tag{S32}$$

and variance

$$\text{Var}\left[k_i^- | \kappa_i^-\right] \simeq \frac{2\pi\mu\langle\kappa\rangle\kappa_i^-}{\beta\sin(\pi/\beta)} = \kappa_i^- \ , \tag{S33}$$

as well as an expression similar to Eq. (S28) for the ensemble average in-degree

$$\langle k^-\rangle \simeq \frac{2\pi\mu\langle\kappa\rangle^2}{\beta\sin(\pi/\beta)} = \langle\kappa\rangle \ . \tag{S34}$$

We also find that the probability for node $i$ with hidden in-degree $\kappa_i^-$ to have a degree equal to $k_i^-$ to be

$$P_i^-(k_i^- | \kappa_i^-) \simeq \frac{[\kappa_i^-]^{k_i^-} e^{-\kappa_i^-}}{k_i^-!} \ , \tag{S35}$$

similarly to Eq. (S30).

### D. Joint in-/out-degree distribution

Since the existence of links is conditionally independent given the values of the hidden in- and out-degrees, the joint in-/out-degree distribution is

$$\begin{aligned} P(k^-, k^+) &= \iint P^-(k^-|\kappa^-)P^+(k^+|\kappa^+)\rho(\kappa^-, \kappa^+)d\kappa^- d\kappa^+ \\ &\simeq \iint \frac{[\kappa^-]^{k^-} e^{-\kappa^-}}{k^-!} \frac{[\kappa^+]^{k^+} e^{-\kappa^+}}{k^+!} \rho(\kappa^-, \kappa^+) d\kappa^- d\kappa^+ \ . \end{aligned} \tag{S36}$$

Hence, the in-degree and out-degree distributions are prescribed by their corresponding marginal pdf of $\rho(\kappa^-, \kappa^+)$ as

$$P^-(k^-) = \sum_{k^+=0}^{\infty} P(k^-, k^+) \simeq \int \frac{e^{-\kappa^-}[\kappa^-]^{k^-}}{k^-!} \left[\int \rho(\kappa^-, \kappa^+) d\kappa^+\right] d\kappa^- \tag{S37a}$$

$$P^+(k^+) = \sum_{k^-=0}^{\infty} P(k^-, k^+) \simeq \int \frac{e^{-\kappa^+}[\kappa^+]^{k^+}}{k^+!} \left[\int \rho(\kappa^-, \kappa^+) d\kappa^-\right] d\kappa^+ \ , \tag{S37b}$$

and the correlations between $k^-$ and $k^+$ are governed by the correlations between $\kappa^-$ and $\kappa^+$ encoded in $\rho(k^-, \kappa^+)$.

### E. Reciprocal degree of nodes

Let us denote the probability for a reciprocal link to exist between nodes $i$ and $j$ by

$$P(a_{ij}=1, a_{ji}=1 | \kappa_i^-, \kappa_i^+, \kappa_j^-, \kappa_j^+, \Delta\theta_{ij}) \ , \tag{S38}$$

with the additional assumption that this connection probability is symmetrical, i.e.

$$P(a_{ij}=1, a_{ji}=1 | \kappa_i^-, \kappa_i^+, \kappa_j^-, \kappa_j^+, \Delta\theta_{ij}) = P(a_{ji}=1, a_{ij}=1 | \kappa_j^-, \kappa_j^+, \kappa_i^-, \kappa_i^+, \Delta\theta_{ji}) \ . \tag{S39}$$

Following similar steps to that in Sec. S.II B, we define the pgf associated with the reciprocal degree of node $i$ given the sequences $\boldsymbol{\theta}$ and $\boldsymbol{\kappa}$ as

$$\begin{aligned} H_i^{\leftrightarrow}(z|\boldsymbol{\kappa}, \boldsymbol{\theta}) &= \sum_{k_i^{\leftrightarrow}=0}^{N-1} P_i^{\leftrightarrow}(k_i^{\leftrightarrow}|\boldsymbol{\theta}, \boldsymbol{\kappa}) z^{k_i^{\leftrightarrow}} \\ &= \prod_{\substack{j=1 \\ j\neq i}}^{N} \left[1 - P(a_{ij}=1, a_{ji}=1|\kappa_i^-, \kappa_i^+, \kappa_j^-, \kappa_j^+, \Delta\theta_{ij}) + zP(a_{ij}=1, a_{ji}=1|\kappa_i^-, \kappa_i^+, \kappa_j^-, \kappa_j^+, \Delta\theta_{ij})\right] \ . \end{aligned} \tag{S40}$$

Hence, general expressions for the expected reciprocal degree of node $i$ and for the ensemble average reciprocal degree are respectively

$$\langle k_i^{\leftrightarrow}|\boldsymbol{\kappa},\boldsymbol{\theta}\rangle = \left.\frac{\partial H_i^{\leftrightarrow}(z|\boldsymbol{\kappa},\boldsymbol{\theta})}{\partial z}\right|_{z=1} = \sum_{\substack{j=1 \\ j\neq i}}^{N} P(a_{ij}=1, a_{ji}=1|\kappa_i^-, \kappa_i^+, \kappa_j^-, \kappa_j^+, \Delta\theta_{ij}),  \tag{S41}$$

and

$$\begin{aligned}\langle k^{\leftrightarrow}|\boldsymbol{\kappa},\boldsymbol{\theta}\rangle &= \frac{1}{N}\sum_{i=1}^{N}\langle k_i^{\leftrightarrow}|\boldsymbol{\kappa},\boldsymbol{\theta}\rangle \\ &= \frac{1}{N}\sum_{i=1}^{N}\sum_{\substack{j=1 \\ j\neq i}}^{N} P(a_{ij}=1, a_{ji}=1|\kappa_i^-,\kappa_i^+,\kappa_j^-,\kappa_j^+,\Delta\theta_{ij}) \\ &= \frac{2}{N}\sum_{i=1}^{N}\sum_{j=i+1}^{N} P(a_{ij}=1, a_{ji}=1|\kappa_i^-,\kappa_i^+,\kappa_j^-,\kappa_j^+,\Delta\theta_{ij}). \end{aligned} \tag{S42}$$

Averaging $H_i^{\leftrightarrow}(z|\boldsymbol{\kappa},\boldsymbol{\theta})$ over every possible sequence $\boldsymbol{\theta}$ and $\boldsymbol{\kappa}\setminus\{\kappa_i^-,\kappa_i^+\}$ yields

$$\begin{aligned} H_i^{\leftrightarrow}(z|\kappa_i^-,\kappa_i^+) &= \int\cdots\int_0^{\pi} H_i^{\leftrightarrow}(z|\boldsymbol{\kappa},\boldsymbol{\theta})\prod_{\substack{j=1 \\ j\neq i}}^{N}\frac{1}{\pi}d\Delta\theta_{ij}\rho(\kappa_j^-,\kappa_j^+)d\kappa_j^-d\kappa_j^+ \\ &= \prod_{\substack{j=1 \\ j\neq i}}^{N}\left[\iiint_0^{\pi}\left[1 - P(a_{ij}=1,a_{ji}=1|\kappa_i^-,\kappa_i^+,\kappa_j^-,\kappa_j^+,\Delta\theta_{ij})\right.\right. \\ &\qquad\left.\left. + zP(a_{ij}=1,a_{ji}=1|\kappa_i^-,\kappa_i^+,\kappa_j^-,\kappa_j^+,\Delta\theta_{ij})\right]\frac{1}{\pi}d\Delta\theta_{ij}\rho(\kappa_j^-,\kappa_j^+)d\kappa_j^-d\kappa_j^+\right] \\ &= \prod_{\substack{j=1 \\ j\neq i}}^{N}\left[\iint\left[1 - \langle a_{ij}a_{ji}|\kappa_i^-,\kappa_i^+,\kappa_j^-,\kappa_j^+\rangle\right.\right. \\ &\qquad\left.\left. + z\langle a_{ij}a_{ji}|\kappa_i^-,\kappa_i^+,\kappa_j^-,\kappa_j^+\rangle\right]\rho(\kappa_j^-,\kappa_j^+)d\kappa_j^-d\kappa_j^+\right] \\ &= \prod_{\substack{j=1 \\ j\neq i}}^{N}\left[1 - \langle a_{i\bullet}a_{\bullet i}|\kappa_i^-,\kappa_i^+\rangle + z\langle a_{i\bullet}a_{\bullet i}|\kappa_i^-,\kappa_i^+\rangle\right] \\ &= \left[1 - \langle a_{i\bullet}a_{\bullet i}|\kappa_i^-,\kappa_i^+\rangle + z\langle a_{i\bullet}a_{\bullet i}|\kappa_i^-,\kappa_i^+\rangle\right]^{N-1}, \end{aligned} \tag{S43}$$

where

$$\langle a_{ij}a_{ji}|\kappa_i^-,\kappa_i^+,\kappa_j^-,\kappa_j^+\rangle = \int_0^{\pi} P(a_{ij}=1,a_{ji}=1|\kappa_i^-,\kappa_i^+,\kappa_j^-,\kappa_j^+,\Delta\theta_{ij})\frac{1}{\pi}d\Delta\theta_{ij} \tag{S44}$$

and

$$\langle a_{i\bullet}a_{\bullet i}|\kappa_i^-,\kappa_i^+\rangle = \iint \langle a_{ij}a_{ji}|\kappa_i^-,\kappa_i^+,\kappa_j^-,\kappa_j^+\rangle\rho(\kappa_j^-,\kappa_j^+)d\kappa_j^-d\kappa_j^+. \tag{S45}$$

The expected reciprocal degree of node $i$ is then

$$\langle k_i^{\leftrightarrow}|\kappa_i^-,\kappa_i^+\rangle = (N-1)\langle a_{i\bullet}a_{\bullet i}|\kappa_i^-,\kappa_i^+\rangle. \tag{S46}$$

The ensemble average expected reciprocal degree is

$$\langle k^{\leftrightarrow}\rangle = \iint \langle k_i^{\leftrightarrow}|\kappa_i^-,\kappa_i^+\rangle\rho(\kappa_i^-,\kappa_i^+)d\kappa_i^-d\kappa_i^+. \tag{S47}$$

## F. Reciprocity

We are now in a position to combine the results from Sec. S.I with those from the previous subsections to study the reciprocity in the networks generated by the directed $\mathbb{S}^1$ model. The reciprocity is defined as

$$r = \frac{L^{\leftrightarrow}}{L} , \tag{S48}$$

where $L$ is the number of links, and $L^{\leftrightarrow}$ is the number of reciprocal links. Note that for $r$ to be such that $0 \leq r \leq 1$, each reciprocal connection (e.g. when two nodes are connected by two links in the opposite direction) must contribute 2 to $L^{\leftrightarrow}$. Hence $L^{\leftrightarrow}$ is an even number. Averaging Eq. (S48) over all possible angular positions $\boldsymbol{\theta}$ and hidden in/out-degrees $\boldsymbol{\kappa}$, we get

$$\langle r \rangle = \left\langle \frac{L^{\leftrightarrow}}{L} \right\rangle \approx \frac{\langle L^{\leftrightarrow}\rangle}{\langle L \rangle} = \frac{N \langle k^{\leftrightarrow}\rangle}{N \langle k^+ \rangle} = \begin{cases} (1+\nu)\langle r|\nu{=}0\rangle - \nu \langle r|\nu{=}{-}1\rangle & -1 \leq \nu \leq 0 \\ \\ (1-\nu)\langle r|\nu{=}0\rangle + \nu \langle r|\nu{=}1\rangle & 0 \leq \nu \leq 1 \end{cases}, \tag{S49}$$

where we used Eqs. (S44)–(S47), and where we defined the following quantities.

1. $\langle r|\nu{=}1\rangle$ is the expected reciprocity when $\nu = 1$

$$\begin{aligned}\langle r|\nu{=}1\rangle &= \frac{\langle k^{\leftrightarrow}|\nu{=}1\rangle}{\langle k^+ \rangle} \\ &= \frac{N-1}{\langle k^+ \rangle} \iiiint \langle a_{ij} a_{ji} | \kappa_i^-, \kappa_i^+, \kappa_j^-, \kappa_j^+, \nu{=}1 \rangle \\ &\qquad\qquad\qquad \times \rho(\kappa_i^-, \kappa_i^+) \rho(\kappa_j^-, \kappa_j^+) d\kappa_i^- d\kappa_i^+ d\kappa_j^- d\kappa_j^+ \end{aligned} \tag{S50}$$

with $\langle a_{ij} a_{ji} | \kappa_i^-, \kappa_i^+, \kappa_j^-, \kappa_j^+, \nu{=}1\rangle$ being the expected reciprocal connection probability, Eq. (S44), when $\nu = 1$

$$\begin{aligned}\langle a_{ij} a_{ji} | \kappa_i^-, \kappa_i^+, \kappa_j^-, \kappa_j^+, \nu{=}1\rangle &= \frac{1}{\pi} \int_0^{\pi} \min\left\{ P(a_{ij}=1|\kappa_i^+, \kappa_j^-, \Delta\theta_{ij}), P(a_{ji}=1|\kappa_j^+, \kappa_i^-, \Delta\theta_{ij}) \right\} d\Delta\theta_{ij} \\ &= \frac{1}{\pi} \int_0^{\pi} \min\left\{ \frac{1}{1+\chi_{ij}^{\beta}}, \frac{1}{1+\chi_{ji}^{\beta}} \right\} d\Delta\theta_{ij} \\ &= H(1-\xi_{ij}) \frac{1}{\pi} \int_0^{\pi} \frac{1}{1+\chi_{ij}^{\beta}} d\Delta\theta_{ij} + H(\xi_{ij}-1) \frac{1}{\pi} \int_0^{\pi} \frac{1}{1+\chi_{ji}^{\beta}} d\Delta\theta_{ij} \\ &= H(1-\xi_{ij}) \frac{2\mu\kappa_i^+ \kappa_j^-}{N} \int_0^{\frac{N}{2\mu\kappa_i^+\kappa_j^-}} \frac{1}{1+\chi_{ij}^{\beta}} d\chi_{ij} \\ &\qquad + H(\xi_{ij}-1) \frac{2\mu\kappa_j^+ \kappa_i^-}{N} \int_0^{\frac{N}{2\mu\kappa_j^+\kappa_i^-}} \frac{1}{1+\chi_{ji}^{\beta}} d\chi_{ij} \\ &= H(1-\xi_{ij}) {}_2F_1\left(1, \frac{1}{\beta}; 1+\frac{1}{\beta}; -\left[\frac{N}{2\mu\kappa_i^+\kappa_j^-}\right]^{\beta}\right) \\ &\qquad + H(\xi_{ij}-1) {}_2F_1\left(1, \frac{1}{\beta}; 1+\frac{1}{\beta}; -\left[\frac{N}{2\mu\kappa_j^+\kappa_i^-}\right]^{\beta}\right) \tag{S51} \\ &\simeq H(1-\xi_{ij}) \frac{2\pi\mu\kappa_i^+\kappa_j^-}{\beta N \sin(\pi/\beta)} + H(\xi_{ij}-1) \frac{2\pi\mu\kappa_j^+\kappa_i^-}{\beta N \sin(\pi/\beta)} \\ &= H(1-\xi_{ij}) \frac{\kappa_i^+ \kappa_j^-}{N\langle\kappa\rangle} + H(\xi_{ij}-1) \frac{\kappa_j^+ \kappa_i^-}{N\langle\kappa\rangle} , \tag{S52}\end{aligned}$$

where we used Eqs. (S95) and (S104), and where we set $\mu = \frac{\beta \sin(\pi/\beta)}{2\pi\langle\kappa\rangle}$ and defined $\xi_{ij} = \frac{\kappa_i^+}{\kappa_i^-} \frac{\kappa_j^-}{\kappa_j^+}$.

2. $\langle r|\nu=0\rangle$ is the expected reciprocity when $\nu = 0$

$$\langle r|\nu=0\rangle = \frac{\langle k^{\leftrightarrow}|\nu=0\rangle}{\langle k^+\rangle}$$

$$= \frac{N-1}{\langle k^+\rangle} \iiiint \langle a_{ij}a_{ji}|\kappa_i^-, \kappa_i^+, \kappa_j^-, \kappa_j^+, \nu=0\rangle$$
$$\times \rho(\kappa_i^-, \kappa_i^+)\rho(\kappa_j^-, \kappa_j^+) d\kappa_i^- d\kappa_i^+ d\kappa_j^- d\kappa_j^+ \quad (S53)$$

with

$$\langle a_{ij}a_{ji}|\kappa_i^-, \kappa_i^+, \kappa_j^-, \kappa_j^+, \nu=0, \xi_{ij}=1\rangle = \frac{1}{\pi}\int_0^\pi P(a_{ij}=1|\kappa_i^+, \kappa_j^-, \Delta\theta_{ij})P(a_{ji}=1|\kappa_j^+, \kappa_i^-, \Delta\theta_{ij})d\Delta\theta_{ij}$$

$$= \frac{1}{\pi}\int_0^\pi \frac{1}{1+\chi_{ij}^\beta}\frac{1}{1+\chi_{ji}^\beta}d\Delta\theta_{ij}$$

$$= \frac{1}{\pi}\int_0^\pi \frac{1}{\left(1+\chi_{ij}^\beta\right)^2}d\Delta\theta_{ij}$$

$$= \frac{2\mu\kappa_i^+\kappa_j^-}{N}\int_0^{\frac{N}{2\mu\kappa_i^+\kappa_j^-}} \frac{1}{\left(1+\chi_{ij}^\beta\right)^2}d\chi_{ij}$$

$$= {}_2F_1\left(2, \frac{1}{\beta}; 1+\frac{1}{\beta}; -\left[\frac{N}{2\mu\kappa_i^+\kappa_j^-}\right]^\beta\right) \quad (S54)$$

$$\simeq \frac{2\pi\mu\kappa_i^+\kappa_j^-}{\beta N \sin(\pi/\beta)}\left(1-\frac{1}{\beta}\right)$$

$$= \frac{\kappa_i^+\kappa_j^-}{N\langle\kappa\rangle}\left(1-\frac{1}{\beta}\right) \quad (S55)$$

when $\xi_{ij} = \frac{\kappa_i^+}{\kappa_i^-}\frac{\kappa_j^-}{\kappa_j^+} = 1$, and

$$\langle a_{ij}a_{ji}|\kappa_i^-, \kappa_i^+, \kappa_j^-, \kappa_j^+, \nu=0, \xi_{ij}\neq 1\rangle = \frac{1}{\pi}\int_0^\pi P(a_{ij}=1|\kappa_i^+, \kappa_j^-, \Delta\theta_{ij})P(a_{ji}=1|\kappa_j^+, \kappa_i^-, \Delta\theta_{ij})d\Delta\theta_{ij}$$

$$= \frac{1}{\pi}\int_0^\pi \frac{1}{1+\chi_{ij}^\beta}\frac{1}{1+\chi_{ji}^\beta}d\Delta\theta_{ij}$$

$$= \frac{1}{\pi}\int_0^\pi \frac{1}{1+\chi_{ij}^\beta}\frac{1}{1+\xi_{ij}^\beta\chi_{ij}^\beta}d\Delta\theta_{ij}$$

$$= \frac{2\mu\kappa_i^+\kappa_j^-}{N}\int_0^{\frac{N}{2\mu\kappa_i^+\kappa_j^-}} \frac{1}{1+\chi_{ij}^\beta}\frac{1}{1+\xi_{ij}^\beta\chi_{ij}^\beta}d\chi_{ij}$$

$$= \frac{1}{1-\xi_{ij}^\beta}{}_2F_1\left(1, \frac{1}{\beta}; 1+\frac{1}{\beta}; -\left[\frac{N}{2\mu\kappa_i^+\kappa_j^-}\right]^\beta\right)$$

$$-\frac{\xi_{ij}^\beta}{1-\xi_{ij}^\beta}{}_2F_1\left(1, \frac{1}{\beta}; 1+\frac{1}{\beta}; -\left[\frac{N}{2\mu\kappa_j^+\kappa_i^-}\right]^\beta\right) \quad (S56)$$

$$\simeq \frac{2\pi\mu\kappa_i^+\kappa_j^-}{\beta N \sin(\pi/\beta)}\frac{1-\xi_{ij}^{\beta-1}}{1-\xi_{ij}^\beta}$$

$$= \frac{\kappa_i^+\kappa_j^-}{N\langle\kappa\rangle}\frac{1-\xi_{ij}^{\beta-1}}{1-\xi_{ij}^\beta} \quad (S57)$$

otherwise. In the last two equations, we again set $\mu = \frac{\beta \sin(\pi/\beta)}{2\pi\langle\kappa\rangle}$, and used Eqs. (S101), (S104) and (S105).

3. $\langle r|\nu{=}{-}1\rangle$ is the expected reciprocity when $\nu = -1$

$$\begin{aligned}\langle r|\nu{=}{-}1\rangle &= \frac{\langle k^{\leftrightarrow}|\nu{=}{-}1\rangle}{\langle k^{+}\rangle} \\ &= \frac{N-1}{\langle k^{+}\rangle}\iiiint \langle a_{ij}a_{ji}|\kappa_i^-,\kappa_i^+,\kappa_j^-,\kappa_j^+,\nu{=}{-}1\rangle \\ &\qquad\qquad\times \rho(\kappa_i^-,\kappa_i^+)\rho(\kappa_j^-,\kappa_j^+)d\kappa_i^- d\kappa_i^+ d\kappa_j^- d\kappa_j^+\end{aligned} \qquad (S58)$$

with

$$\begin{aligned}\langle a_{ij}a_{ji}|\kappa_i^-,\kappa_i^+,\kappa_j^-,\kappa_j^+,\nu=-1\rangle &= \frac{1}{\pi}\int_0^{\pi}\Big[P(a_{ij}=1|\kappa_i^+,\kappa_j^-,\Delta\theta_{ij}) + P(a_{ji}=1|\kappa_j^+,\kappa_i^-,\Delta\theta_{ij}) - 1\Big] \\ &\qquad\times H\Big(P(a_{ij}=1|\kappa_i^+,\kappa_j^-,\Delta\theta_{ij}) + P(a_{ji}=1|\kappa_j^+,\kappa_i^-,\Delta\theta_{ij}) - 1\Big)d\Delta\theta_{ij} \\ &= \frac{1}{\pi}\int_0^{\Delta\theta_{ij}^c}\Big[P(a_{ij}=1|\kappa_i^+,\kappa_j^-,\Delta\theta_{ij}) + P(a_{ji}=1|\kappa_j^+,\kappa_i^-,\Delta\theta_{ij}) - 1\Big]d\Delta\theta_{ij} \\ &= \frac{1}{\pi}\int_0^{\Delta\theta_{ij}^c}\frac{d\Delta\theta_{ij}}{1+\chi_{ij}^{\beta}} + \frac{1}{\pi}\int_0^{\Delta\theta_{ij}^c}\frac{d\Delta\theta_{ij}}{1+\chi_{ji}^{\beta}} - \frac{1}{\pi}\int_0^{\Delta\theta_{ij}^c}d\Delta\theta_{ij} \\ &= \frac{2\mu\kappa_i^+\kappa_j^-}{N}\int_0^{\frac{N\Delta\theta_{ij}^c}{2\pi\mu\kappa_i^+\kappa_j^-}}\frac{d\chi_{ij}}{1+\chi_{ij}^{\beta}} + \frac{2\mu\kappa_j^+\kappa_i^-}{N}\int_0^{\frac{N\Delta\theta_{ij}^c}{2\pi\mu\kappa_j^+\kappa_i^-}}\frac{d\chi_{ji}}{1+\chi_{ji}^{\beta}} - \frac{\Delta\theta_{ij}^c}{\pi} \\ &= \frac{\Delta\theta_{ij}^c}{\pi}\Bigg[{}_2F_1\Bigg(1,\frac{1}{\beta};1+\frac{1}{\beta};-\Bigg[\frac{N\Delta\theta_{ij}^c}{2\pi\mu\kappa_i^+\kappa_j^-}\Bigg]^{\beta}\Bigg) \\ &\qquad + {}_2F_1\Bigg(1,\frac{1}{\beta};1+\frac{1}{\beta};-\Bigg[\frac{N\Delta\theta_{ij}^c}{2\pi\mu\kappa_j^+\kappa_i^-}\Bigg]^{\beta}\Bigg) - 1\Bigg]\end{aligned} \qquad (S59)$$

where we used Eq. (S95), and where $\Delta\theta_{ij}^c$ is the solution of

$$P(a_{ij}=1|\kappa_i^+,\kappa_j^-,\Delta\theta_{ij}^c) + P(a_{ji}=1|\kappa_j^+,\kappa_i^-,\Delta\theta_{ij}^c) = 1 . \qquad (S60)$$

To explore the limit $N\to\infty$ such that $N/(\kappa_i^+\kappa_j^-)\to\infty$ and $N/(\kappa_j^+\kappa_i^-)\to\infty$, we note that

$$\begin{aligned}1 &= P(a_{ij}=1|\kappa_i^+,\kappa_j^-,\Delta\theta_{ij}^c) + P(a_{ji}=1|\kappa_j^+,\kappa_i^-,\Delta\theta_{ij}^c) \\ &= \frac{1}{1+\Big[\frac{N\Delta\theta_{ij}^c}{2\pi\mu\kappa_i^+\kappa_j^-}\Big]^{\beta}} + \frac{1}{1+\Big[\frac{N\Delta\theta_{ij}^c}{2\pi\mu\kappa_j^+\kappa_i^-}\Big]^{\beta}} \\ &\simeq \frac{[2\pi\mu\kappa_i^+\kappa_j^-]^{\beta}}{[N\Delta\theta_{ij}^c]^{\beta}} + \frac{[2\pi\mu\kappa_j^+\kappa_i^-]^{\beta}}{[N\Delta\theta_{ij}^c]^{\beta}} ,\end{aligned} \qquad (S61)$$

and thus

$$\Delta\theta_{ij}^c \simeq \frac{2\pi\mu}{N}\Big[[\kappa_i^+\kappa_j^-]^{\beta} + [\kappa_j^+\kappa_i^-]^{\beta}\Big]^{\frac{1}{\beta}} = \frac{2\pi\mu\kappa_i^+\kappa_j^-}{N}\Big[1+\xi_{ij}^{-\beta}\Big]^{\frac{1}{\beta}} . \qquad (S62)$$

Equation (S59) then becomes

$$\langle a_{ij}a_{ji}|\kappa_i^-, \kappa_i^+, \kappa_j^-, \kappa_j^+, \nu = -1\rangle \simeq \frac{2\mu\kappa_i^+\kappa_j^-}{N}\left[1+\xi_{ij}^{-\beta}\right]^{\frac{1}{\beta}}\left[{}_2F_1\left(1,\frac{1}{\beta};1+\frac{1}{\beta};-1-\xi_{ij}^{-\beta}\right)\right.$$

$$\left.+{}_2F_1\left(1,\frac{1}{\beta};1+\frac{1}{\beta};-1-\xi_{ij}^{\beta}\right)-1\right]$$

$$=\frac{\kappa_i^+\kappa_j^-}{N\langle\kappa\rangle}\frac{\sin(\pi/\beta)}{\pi/\beta}\left[1+\xi_{ij}^{-\beta}\right]^{\frac{1}{\beta}}\left[{}_2F_1\left(1,\frac{1}{\beta};1+\frac{1}{\beta};-1-\xi_{ij}^{-\beta}\right)\right.$$

$$\left.+{}_2F_1\left(1,\frac{1}{\beta};1+\frac{1}{\beta};-1-\xi_{ij}^{\beta}\right)-1\right], \quad\text{(S63)}$$

where we set $\mu = \dfrac{\beta\sin(\pi/\beta)}{2\pi\langle\kappa\rangle}$.

## S.III. NETWORK DATASETS

The network datasets used in the article have been made publicly available by the original authors and were downloaded from The Netzschleuder network catalogue and repository (https://networks.skewed.de). The name of each dataset is listed below:

7th_graders, add_health_comm15, add_health_comm27, add_health_comm28, add_health_comm33, add_health_comm40, add_health_comm41, add_health_comm50, add_health_comm61, add_health_comm62, add_health_comm68, add_health_comm73, add_health_comm75, add_health_comm79, add_health_comm81, add_health_comm83, add_health_comm84, advogato, anybeat, bison, bitcoin_alpha, bitcoin_trust, caida_as_20040105, caida_as_20040202, caida_as_20040301, caida_as_20040405, caida_as_20040503, caida_as_20040607, caida_as_20040705, caida_as_20040802, caida_as_20040906, caida_as_20041004, caida_as_20041101, caida_as_20041206, caida_as_20050103, caida_as_20050207, caida_as_20050307, caida_as_20050404, caida_as_20050502, caida_as_20050606, caida_as_20050704, caida_as_20050801, caida_as_20050905, caida_as_20051003, caida_as_20051107, caida_as_20051205, caida_as_20060102, caida_as_20060109, caida_as_20060116, caida_as_20060123, caida_as_20060130, caida_as_20060206, caida_as_20060213, caida_as_20060220, caida_as_20060227, caida_as_20060306, caida_as_20060313, caida_as_20060320, caida_as_20060327, caida_as_20060403, caida_as_20060410, caida_as_20060417, caida_as_20060424, caida_as_20060501, caida_as_20060508, caida_as_20060515, caida_as_20060522, caida_as_20060529, caida_as_20060605, caida_as_20060612, caida_as_20060619, caida_as_20060626, caida_as_20060703, caida_as_20060710, caida_as_20060717, caida_as_20060724, caida_as_20060731, caida_as_20060807, caida_as_20060814, caida_as_20060821, caida_as_20060828, caida_as_20060904, caida_as_20060911, caida_as_20060918, caida_as_20060925, caida_as_20061002, caida_as_20061009, caida_as_20061016, caida_as_20061023, caida_as_20061030, caida_as_20061106, caida_as_20061113, caida_as_20061120, caida_as_20061127, caida_as_20061204, caida_as_20061211, caida_as_20061218, caida_as_20061225, caida_as_20070101, caida_as_20070108, caida_as_20070115, caida_as_20070122, caida_as_20070129, caida_as_20070205, caida_as_20070212, caida_as_20070219, caida_as_20070226, caida_as_20070305, caida_as_20070312, caida_as_20070423, caida_as_20070917, cattle, celegans_2019_hermaphrodite_chemical, celegans_2019_hermaphrodite_chemical_corrected, celegans_2019_hermaphrodite_chemical_synapse, celegans_2019_male_chemical, celegans_2019_male_chemical_corrected, celegans_2019_male_chemical_synapse, celegansneural, chess, chicago_road, cintestinalis, college_freshmen, copenhagen_calls, copenhagen_sms, cora, dblp_cite, dutch_school_klas12b-net-1, dutch_school_klas12b-net-2, dutch_school_klas12b-net-3, dutch_school_klas12b-net-3m, dutch_school_klas12b-net-4, dutch_school_klas12b-net-4m, dutch_school_klas12b-primary, ecoli_transcription_v1_0, ecoli_transcription_v1_1, email_company, faa_routes, fao_trade, foodweb_baywet, foodweb_little_rock, fresh_webs_AkatoreA, fresh_webs_AkatoreB, fresh_webs_Berwick, fresh_webs_Blackrock, fresh_webs_Broad, fresh_webs_Canton, fresh_webs_Catlins, fresh_webs_Coweeta1, fresh_webs_Coweeta17, fresh_webs_DempstersAu, fresh_webs_DempstersSp, fresh_webs_DempstersSu, fresh_webs_German, fresh_webs_Healy, fresh_webs_Kyeburn, fresh_webs_LilKyeburn, fresh_webs_Martins, fresh_webs_Narrowdale, fresh_webs_NorthCol, fresh_webs_Powder, fresh_webs_Stony, fresh_webs_SuttonAu, fresh_webs_SuttonSp, fresh_webs_SuttonSu, fresh_webs_Troy, fresh_webs_Venlaw, freshman_t0, freshman_t2, freshman_t3, freshman_t5

| | | |
|---|---|---|
| freshman_t6 | freshmen_t0 | freshmen_t2 |
| freshmen_t3 | freshmen_t5 | freshmen_t6 |
| genetic_multiplex_Arabidopsis | genetic_multiplex_Bos_Multiplex_Genetic | genetic_multiplex_Candida |
| genetic_multiplex_Celegans | genetic_multiplex_DanioRerio | genetic_multiplex_Drosophila |
| genetic_multiplex_Gallus | genetic_multiplex_HepatitusCVirus | genetic_multiplex_HumanHIV1 |
| genetic_multiplex_HumanHerpes4 | genetic_multiplex_Mus | genetic_multiplex_Oryctolagus |
| genetic_multiplex_Plasmodium | genetic_multiplex_Rattus | genetic_multiplex_Sacchpomb |
| genetic_multiplex_Xenopus | gnutella_04 | gnutella_06 |
| gnutella_08 | gnutella_09 | gnutella_25 |
| hens | high_tech_company | highschool |
| inploid | interactome_figeys | interactome_stelzl |
| jdk | jung | law_firm |
| macaques | messal_shale | moreno_sheep |
| moreno_taro | openflights | packet_delays |
| physician_trust | polblogs | qa_user_mathoverflow_a2q |
| qa_user_mathoverflow_c2a | qa_user_mathoverflow_c2q | residence_hall |
| rhesus_monkey | sp_high_school_diaries | sp_high_school_survey |
| un_migrations | uni_email | us_agencies_alabama |
| us_agencies_alaska | us_agencies_arizona | us_agencies_arkansas |
| us_agencies_california | us_agencies_colorado | us_agencies_connecticut |
| us_agencies_delaware | us_agencies_florida | us_agencies_georgia |
| us_agencies_hawaii | us_agencies_idaho | us_agencies_illinois |
| us_agencies_indiana | us_agencies_iowa | us_agencies_kansas |
| us_agencies_kentucky | us_agencies_louisiana | us_agencies_maine |
| us_agencies_maryland | us_agencies_massachusetts | us_agencies_michigan |
| us_agencies_minnesota | us_agencies_mississippi | us_agencies_missouri |
| us_agencies_montana | us_agencies_nebraska | us_agencies_nevada |
| us_agencies_newhampshire | us_agencies_newjersey | us_agencies_newmexico |
| us_agencies_newyork | us_agencies_northcarolina | us_agencies_northdakota |
| us_agencies_ohio | us_agencies_oklahoma | us_agencies_oregon |
| us_agencies_pennsylvania | us_agencies_rhodeisland | us_agencies_southcarolina |
| us_agencies_southdakota | us_agencies_tennessee | us_agencies_texas |
| us_agencies_utah | us_agencies_vermont | us_agencies_virginia |
| us_agencies_washington | us_agencies_westvirginia | us_agencies_wisconsin |
| us_agencies_wyoming | us_air_traffic | webkb_webkb_cornell_link1 |
| webkb_webkb_texas_link1 | webkb_webkb_washington_link1 | webkb_webkb_wisconsin_link1 |
| wiki_talk_br | wiki_talk_cy | wiki_talk_eo |
| wiki_talk_gl | wiki_talk_ht | wiki_talk_nds |
| wiki_talk_oc | wikipedia_link_si | word_adjacency_darwin |
| word_adjacency_french | word_adjacency_japanese | word_adjacency_spanish |
| yeast_transcription | | |

## S.IV. INFERENCE ALGORITHM

The inference algorithm used in the main text is an adaptation of the parameter inference procedure of the embedding algorithm introduced in Ref. [1]. Its objective is to infer the $2N + 2$ parameters $\boldsymbol{\kappa} = \kappa_1^-, \kappa_1^+, \ldots, \kappa_N^-, \kappa_N^+, \beta$ and $\nu$ so that the directed $\mathbb{S}^1$ model will reproduce, on average, the joint in/out-degree sequence, the reciprocity and the density of triangles of an original real directed network ($2N + 2$ constraints).

Note that, contrary to the embedding algorithm introduced in Ref. [1], the inference algorithm does not aim to infer the angular positions, $\boldsymbol{\theta}$; the aforementioned $2N + 2$ parameters are therefore inferred when averaging over all possible angular positions.

### A. Inputs

The following $2N + 2$ constraints are measured on an original real directed network and used as inputs for the inference algorithm.

1. The joint in/out-degree sequence $\boldsymbol{k} = \{k_1^-, k_1^+, \ldots, k_N^-, k_N^+\}$, where

$$k_i^- = |\partial_i^-| \tag{S64a}$$
$$k_i^+ = |\partial_i^+|, \tag{S64b}$$

and where $\partial_i^-$ ($\partial_i^+$) is the set of in-neighbors (out-neighbors) of node $i$ in the original real directed network.

2. The reciprocity $r^{\text{obs}}$ computed as

$$r^{\text{obs}} = \frac{L^{\leftrightarrow}}{L} = \frac{\sum_{i=1}^N |\partial_i^- \cap \partial_i^+|}{\sum_{i=1}^N |\partial_i^+|}, \tag{S65}$$

where $|\partial_i^- \cap \partial_i^+|$ counts the number of neighbors with which node $i$ shares both possible directed links (i.e. reciprocal connection).

3. The density of triangles, $\bar{c}_{\text{obs}}$, as measured by the average undirected local clustering coefficient

$$\bar{c}_{\text{undir}}^{\text{obs}} = \frac{1}{N_{>1}} \sum_{i=1}^N c_i = \frac{1}{N_{>1}} \sum_{i=1}^N \frac{2T_i}{|\partial_i^- \cup \partial_i^+|(|\partial_i^- \cup \partial_i^+| - 1)} \tag{S66}$$

where $T_i$ is the number of triangles to which node $i$ participates, where the quantity $|\partial_i^- \cup \partial_i^+|$ corresponds to the degree of node $i$ in the undirected version of the network, and where $N_{>1}$ is the number of nodes for which $|\partial_i^- \cup \partial_i^+| > 1$. Note that we set $c_i = 0$ for the $N - N_{>1}$ nodes for which $|\partial_i^- \cup \partial_i^+| < 2$.

### B. Inferring the hidden in/out-degrees

This subroutine assumes that a maximal deviation tolerance $\varepsilon_{\text{tol}}^{\max}$ and the parameter $\beta$ have both been assigned some value (e.g. $\varepsilon_{\text{tol}}^{\max} = 0.01$), and uses

$$\mu = \frac{\beta \sin\left(\frac{\pi}{\beta}\right)}{2\pi \langle k^+ \rangle} \tag{S67}$$

where $\langle k^+ \rangle$ is the average out-degree (or equivalently average in-degree) in the original real directed network

$$\langle k^+ \rangle = \frac{1}{N} \sum_{i=1}^N k_i^+ = \frac{1}{N} \sum_{i=1}^N k_i^-. \tag{S68}$$

1. *Initialize* the hidden in/out-degrees by setting $\kappa_i^- = k_i^-$ and $\kappa_i^+ = k_i^+$ for all $i = 1, \ldots, N$.

2. *Compute expected in/out-degrees* as

$$\langle k_i^-|\boldsymbol{\kappa}\rangle = \sum_{\substack{j=1 \\ j\neq i}}^{N} \langle a_{ji}|\kappa_j^+,\kappa_i^-\rangle = \sum_{\substack{j=1 \\ j\neq i}}^{N} {}_2F_1\left(1,\frac{1}{\beta};1+\frac{1}{\beta};-\left[\frac{N}{2\mu\kappa_j^+\kappa_i^-}\right]^{\beta}\right), \tag{S69a}$$

$$\langle k_i^+|\boldsymbol{\kappa}\rangle = \sum_{\substack{j=1 \\ j\neq i}}^{N} \langle a_{ij}|\kappa_i^+,\kappa_j^-\rangle = \sum_{\substack{j=1 \\ j\neq i}}^{N} {}_2F_1\left(1,\frac{1}{\beta};1+\frac{1}{\beta};-\left[\frac{N}{2\mu\kappa_i^+\kappa_j^-}\right]^{\beta}\right), \tag{S69b}$$

for all $i = 1,\ldots,N$. Equations (S69a) and (S69b) are obtained by averaging Eq. (S20) (and its equivalent for in-degrees) over all angular positions, combined with Eq. (S23a).

3. *Compute the largest deviation*, $\varepsilon^{\max}$, between the expected in/out-degrees and the in/out-degrees in the original network as

$$\varepsilon^{\max} = \max\left\{\max\left\{|\langle k_i^-|\boldsymbol{\kappa}\rangle - k_i^-|,|\langle k_i^+|\boldsymbol{\kappa}\rangle - k_i^+|\right\} : i = 1,\ldots,N\right\}. \tag{S70}$$

The hidden in/out-degrees have converged to acceptable values if $\varepsilon^{\max} < \varepsilon^{\max}_{\text{tol}}$ and we proceed to step 6. Otherwise, they require more refinement and we proceed to step 4.

4. *Update the hidden in/out-degrees* according to

$$\kappa_i^- \leftarrow \left|\kappa_i^- + [k_i^- - \langle k_i^-|\boldsymbol{\kappa}\rangle]u^-\right|, \tag{S71a}$$

$$\kappa_i^+ \leftarrow \left|\kappa_i^+ + [k_i^+ - \langle k_i^+|\boldsymbol{\kappa}\rangle]u^+\right|, \tag{S71b}$$

for all $i = 1,\ldots,N$, and where $u^- \sim \text{Uniform}(0,1)$ and $u^+ \sim \text{Uniform}(0,1)$. The random variables prevent the subroutine from getting trapped in a local minimum.

5. *Proceed to step 2* using the updated values for the hidden in/out-degrees.

6. *Compute the expected in-degree and out-degree* as

$$\langle k^-|\boldsymbol{\kappa}\rangle = \frac{1}{N}\sum_{i=1}^{N}\langle k_i^-|\boldsymbol{\kappa}\rangle \tag{S72a}$$

$$\langle k^+|\boldsymbol{\kappa}\rangle = \frac{1}{N}\sum_{i=1}^{N}\langle k_i^+|\boldsymbol{\kappa}\rangle. \tag{S72b}$$

Equations (S72a) and (S72b) are obtained by averaging Eq. (S22) (and its equivalent for in-degrees) over all angular positions. Note that $\langle k^-|\boldsymbol{\kappa}\rangle$ and $\langle k^+|\boldsymbol{\kappa}\rangle$ will be equal up to the numerical error induced by $\varepsilon^{\max}_{\text{tol}}$.

### C. Inferring parameter $\nu$

This subroutine assumes that the parameter $\beta$ has been assigned some value, and uses the parameter $\mu$, the hidden in/out-degrees, $\boldsymbol{\kappa} = \kappa_1^-,\kappa_1^+,\ldots,\kappa_N^-,\kappa_N^+$, as well as the expected out-degree, $\langle k^+|\boldsymbol{\kappa}\rangle$, computed in Sec. S.IV B.

The expected reciprocity in the directed $\mathbb{S}^1$ model is computed as

$$\langle r|\boldsymbol{\kappa}\rangle = \left\langle\frac{L^{\leftrightarrow}}{L}\bigg|\boldsymbol{\kappa}\right\rangle \approx \frac{\langle L^{\leftrightarrow}|\boldsymbol{\kappa}\rangle}{\langle L|\boldsymbol{\kappa}\rangle} = \frac{N\langle k^{\leftrightarrow}|\boldsymbol{\kappa}\rangle}{N\langle k^+|\boldsymbol{\kappa}\rangle}, \tag{S73}$$

where $\langle k^+|\boldsymbol{\kappa}\rangle$ is taken from Eq. (S72b) and $\langle k^{\leftrightarrow}|\boldsymbol{\kappa}\rangle$ is computed by averaging Eq. (S42) over all angular positions. Equation (S73) then takes a similar form as Eq. (S49) and becomes

$$\langle r|\boldsymbol{\kappa}\rangle \approx \begin{cases} (1+\nu)\langle r|\boldsymbol{\kappa},\nu=0\rangle - \nu\langle r|\boldsymbol{\kappa},\nu=-1\rangle & \text{if } -1 \leq \nu \leq 0 \\ (1-\nu)\langle r|\boldsymbol{\kappa},\nu=0\rangle + \nu\langle r|\boldsymbol{\kappa},\nu=1\rangle & \text{if } 0 \leq \nu \leq 1 \end{cases}, \tag{S74}$$

and the inferred value of $\nu$ is obtained such that $\langle r|\boldsymbol{\kappa}\rangle = r^{\text{obs}}$. This subroutine computes $\langle r|\boldsymbol{\kappa}\rangle$ and $\nu$ via the following steps.

1. *Compute the expected reciprocity when $\nu = 1$ using*

$$\langle r|\boldsymbol{\kappa},\nu{=}1\rangle = \frac{\langle k^{\leftrightarrow}|\boldsymbol{\kappa},\nu{=}1\rangle}{\langle k^+|\boldsymbol{\kappa}\rangle} = \frac{2}{N\langle k^+|\boldsymbol{\kappa}\rangle} \sum_{i=1}^{N}\sum_{j=i+1}^{N} \langle a_{ij}a_{ji}|\kappa_i^-,\kappa_i^+,\kappa_j^-,\kappa_j^+,\nu{=}1\rangle \quad \text{(S75a)}$$

where

$$\langle a_{ij}a_{ji}|\kappa_i^-,\kappa_i^+,\kappa_j^-,\kappa_j^+,\nu{=}1\rangle = \begin{cases} {}_2F_1\left(1,\frac{1}{\beta};1+\frac{1}{\beta};-\left[\frac{N}{2\mu\kappa_i^+\kappa_j^-}\right]^{\beta}\right) & \text{if } \xi_{ij} = \frac{\kappa_i^+}{\kappa_i^-}\frac{\kappa_j^-}{\kappa_j^+} < 1 \\ \\ {}_2F_1\left(1,\frac{1}{\beta};1+\frac{1}{\beta};-\left[\frac{N}{2\mu\kappa_j^+\kappa_i^-}\right]^{\beta}\right) & \text{if } \xi_{ij} = \frac{\kappa_i^+}{\kappa_i^-}\frac{\kappa_j^-}{\kappa_j^+} > 1 \end{cases}. \quad \text{(S75b)}$$

Equations (S75a) and (S75b) are obtained by averaging Eq. (S42) over all angular positions, combined with Eq. (S51).

2. *Compute the expected reciprocity when $\nu = 0$ using*

$$\langle r|\boldsymbol{\kappa},\nu{=}0\rangle = \frac{\langle k^{\leftrightarrow}|\boldsymbol{\kappa},\nu{=}0\rangle}{\langle k^+|\boldsymbol{\kappa}\rangle} = \frac{2}{N\langle k^+|\boldsymbol{\kappa}\rangle} \sum_{i=1}^{N}\sum_{j=i+1}^{N} \langle a_{ij}a_{ji}|\kappa_i^-,\kappa_i^+,\kappa_j^-,\kappa_j^+,\nu{=}0\rangle \quad \text{(S76a)}$$

where

$$\langle a_{ij}a_{ji}|\kappa_i^-,\kappa_i^+,\kappa_j^-,\kappa_j^+,\nu{=}0\rangle = \begin{cases} \frac{1}{1-\xi_{ij}^{\beta}}{}_2F_1\left(1,\frac{1}{\beta};1+\frac{1}{\beta};-\left[\frac{N}{2\mu\kappa_i^+\kappa_j^-}\right]^{\beta}\right) \\ \quad -\frac{\xi_{ij}^{\beta}}{1-\xi_{ij}^{\beta}}{}_2F_1\left(1,\frac{1}{\beta};1+\frac{1}{\beta};-\left[\frac{N}{2\mu\kappa_j^+\kappa_i^-}\right]^{\beta}\right) & \text{if } \xi_{ij} = \frac{\kappa_i^+}{\kappa_i^-}\frac{\kappa_j^-}{\kappa_j^+} \neq 1 \\ \\ {}_2F_1\left(2,\frac{1}{\beta};1+\frac{1}{\beta};-\left[\frac{N}{2\mu\kappa_i^+\kappa_j^-}\right]^{\beta}\right) & \text{if } \xi_{ij} = \frac{\kappa_i^+}{\kappa_i^-}\frac{\kappa_j^-}{\kappa_j^+} = 1 \end{cases}. \quad \text{(S76b)}$$

Equations (S76a) and (S76b) are obtained by averaging Eq. (S42) over all angular positions, combined with Eqs. (S54) and (S56).

3. *Compute the expected reciprocity when $\nu = -1$ using*

$$\langle r|\boldsymbol{\kappa},\nu{=}0\rangle = \frac{\langle k^{\leftrightarrow}|\boldsymbol{\kappa},\nu{=}-1\rangle}{\langle k^+|\boldsymbol{\kappa}\rangle} = \frac{2}{N\langle k^+|\boldsymbol{\kappa}\rangle} \sum_{i=1}^{N}\sum_{j=i+1}^{N} \langle a_{ij}a_{ji}|\kappa_i^-,\kappa_i^+,\kappa_j^-,\kappa_j^+,\nu{=}-1\rangle \quad \text{(S77a)}$$

where

$$\langle a_{ij}a_{ji}|\kappa_i^-,\kappa_i^+,\kappa_j^-,\kappa_j^+,\nu{=}-1\rangle = \frac{\Delta\theta_{ij}^{\text{c}}}{\pi}\left[{}_2F_1\left(1,\frac{1}{\beta};1+\frac{1}{\beta};-\left[\frac{N\Delta\theta_{ij}^{\text{c}}}{2\pi\mu\kappa_i^+\kappa_j^-}\right]^{\beta}\right)\right.$$
$$\left. + {}_2F_1\left(1,\frac{1}{\beta};1+\frac{1}{\beta};-\left[\frac{N\Delta\theta_{ij}^{\text{c}}}{2\pi\mu\kappa_j^+\kappa_i^-}\right]^{\beta}\right) - 1\right] \quad \text{(S77b)}$$

and where $\Delta\theta_{ij}^{\text{c}} \in [0,\pi]$ is the solution of

$$P(a_{ij} = 1|\kappa_i^+,\kappa_j^-,\Delta\theta_{ij}^{\text{c}}) + P(a_{ji} = 1|\kappa_j^+,\kappa_i^-,\Delta\theta_{ij}^{\text{c}}) = 1 . \quad \text{(S77c)}$$

Equations (S77a)–(S77c) are obtained by averaging Eq. (S42) over all angular positions, combined with Eq. (S59).

4. *Compute the inferred value of $\nu$ according to*

$$\nu = \begin{cases} \dfrac{r^{\mathrm{obs}} - \langle r|\boldsymbol{\kappa}, \nu=0\rangle}{\langle r|\boldsymbol{\kappa}, \nu=-1\rangle + \langle r|\boldsymbol{\kappa}, \nu=0\rangle} & \text{if } r^{\mathrm{obs}} < \langle r|\boldsymbol{\kappa}, \nu=0\rangle \\[2ex] \dfrac{r^{\mathrm{obs}} - \langle r|\boldsymbol{\kappa}, \nu=0\rangle}{\langle r|\boldsymbol{\kappa}, \nu=1\rangle - \langle r|\boldsymbol{\kappa}, \nu=0\rangle} & \text{if } r^{\mathrm{obs}} > \langle r|\boldsymbol{\kappa}, \nu=0\rangle \end{cases}. \tag{S78}$$

### D. Estimating the expected density of triangles

This subroutine assumes that the parameter $\beta$ has been assigned some value, uses the parameter $\nu$ computed in Sec. S.IV C, and uses the parameter $\mu$ as well as the hidden in/out-degrees, $\boldsymbol{\kappa} = \kappa_1^-, \kappa_1^+, \ldots, \kappa_N^-, \kappa_N^+$ computed in Sec. S.IV B.

The density of triangles is quantified using the average undirected local clustering coefficient, that is the average local clustering coefficient measured on the *undirected projection* of the directed network. This projection is specified via its adjacency matrix, $\tilde{\mathbf{A}}$, whose elements are $\tilde{a}_{ij} = \max(a_{ij}, a_{ji})$. In other words, two nodes are connected in the projection if they are connected by at least one directed link, which occurs with probability

$$\begin{aligned} P(\tilde{a}_{ij} = 1|\kappa_i^-, \kappa_i^+, \kappa_j^-, \kappa_j^+, \Delta\theta_{ij}) &= P_{ij}(a_{ij} = 1, a_{ji} = 0|\kappa_i^-, \kappa_i^+, \kappa_j^-, \kappa_j^+, \Delta\theta_{ij}) \\ &+ P_{ij}(a_{ij} = 0, a_{ji} = 1|\kappa_i^-, \kappa_i^+, \kappa_j^-, \kappa_j^+, \Delta\theta_{ij}) \\ &+ P_{ij}(a_{ij} = 1, a_{ji} = 1|\kappa_i^-, \kappa_i^+, \kappa_j^-, \kappa_j^+, \Delta\theta_{ij}). \end{aligned} \tag{S79}$$

This last expression can be rewritten as

$$\begin{aligned} P(\tilde{a}_{ij} = 1|\kappa_i^-, \kappa_i^+, \kappa_j^-, \kappa_j^+, \Delta\theta_{ij}) &= P_{ij}(a_{ij} = 1|\kappa_i^+, \kappa_j^-, \Delta\theta_{ij}) \\ &+ P_{ij}(a_{ji} = 1|\kappa_i^-, \kappa_j^+, \Delta\theta_{ij}) \\ &- P_{ij}(a_{ij} = 1, a_{ji} = 1|\kappa_i^-, \kappa_i^+, \kappa_j^-, \kappa_j^+, \Delta\theta_{ij}), \end{aligned} \tag{S80}$$

where the three probabilities on the right-hand side are obtained using Eqs. (S15) and (S18). Let us also introduce $P(\tilde{a}_{ij} = 1|\kappa_i^-, \kappa_i^+, \kappa_j^-, \kappa_j^+)$ which corresponds to $P(\tilde{a}_{ij} = 1|\kappa_i^-, \kappa_i^+, \kappa_j^-, \kappa_j^+, \Delta\theta_{ij})$ averaged over all possible angular positions

$$\begin{aligned} P(\tilde{a}_{ij} = 1|\kappa_i^-, \kappa_i^+, \kappa_j^-, \kappa_j^+) &= \int P(\tilde{a}_{ij} = 1|\kappa_i^-, \kappa_i^+, \kappa_j^-, \kappa_j^+) P(\Delta\theta_{ij}) d\Delta\theta_{ij} \\ &= \langle a_{ij}|\kappa_i^+, \kappa_j^-\rangle + \langle a_{ji}|\kappa_j^+, \kappa_i^-\rangle - \langle a_{ij}a_{ji}|\kappa_i^-, \kappa_i^+, \kappa_j^-, \kappa_j^+\rangle, \end{aligned} \tag{S81}$$

where $\langle a_{ij}|\kappa_i^+, \kappa_j^-\rangle$ and $\langle a_{ji}|\kappa_j^+, \kappa_i^-\rangle$ are computed using Eq. (S23a), and where

$$\langle a_{ij}a_{ji}|\kappa_i^-, \kappa_i^+, \kappa_j^-, \kappa_j^+\rangle = \begin{cases} (1+\nu)\langle a_{ij}a_{ji}|\kappa_i^-, \kappa_i^+, \kappa_j^-, \kappa_j^+, \nu=0\rangle \\ \qquad -\nu\langle a_{ij}a_{ji}|\kappa_i^-, \kappa_i^+, \kappa_j^-, \kappa_j^+, \nu=-1\rangle & -1 \leq \nu \leq 0 \\[2ex] (1-\nu)\langle a_{ij}a_{ji}|\kappa_i^-, \kappa_i^+, \kappa_j^-, \kappa_j^+, \nu=0\rangle \\ \qquad +\nu\langle a_{ij}a_{ji}|\kappa_i^-, \kappa_i^+, \kappa_j^-, \kappa_j^+, \nu=1\rangle & 0 \leq \nu \leq 1 \end{cases} \tag{S82}$$

is computed using Eqs. (S51), (S54), (S56) and (S59).

From these quantities, we use Bayes theorem to define two probability distributions with which we estimate the expected density of triangles. The first one corresponds to the probability that neighbor $j$ of node $i$ has hidden degrees $\kappa_j^-, \kappa_j^+$ regardless of the angular distance

$$P(\kappa_j^-, \kappa_j^+|\tilde{a}_{ij} = 1, \kappa_i^-, \kappa_i^+) = \frac{P(\tilde{a}_{ij} = 1|\kappa_i^-, \kappa_i^+, \kappa_j^-, \kappa_j^+)P(\kappa_j^-, \kappa_j^+)}{P(\tilde{a}_{i\bullet} = 1|\kappa_i^-, \kappa_i^+)} \tag{S83}$$

where $P(\tilde{a}_{i\bullet} = 1|\kappa_i^-, \kappa_i^+)$ is a normalization constant. The second distribution provides the probability that neighboring nodes $i$ and $j$ are at angular distance $\Delta\theta_{ij}$

$$P(\Delta\theta_{ij}|\tilde{a}_{ij} = 1, \kappa_i^-, \kappa_i^+, \kappa_j^-, \kappa_j^+) = \frac{P(\tilde{a}_{ij} = 1|\kappa_i^-, \kappa_i^+, \kappa_j^-, \kappa_j^+, \Delta\theta_{ij})P(\Delta\theta_{ij})}{P(\tilde{a}_{ij} = 1|\kappa_i^-, \kappa_i^+, \kappa_j^-, \kappa_j^+)}. \tag{S84}$$

Recall that $P(\Delta\theta_{ij}) = 1/\pi$ in the directed $\mathbb{S}^1$ model.

With these quantities in hand, the expected density of triangles is estimated by computing

$$\bar{c}_{\text{undir}} \approx \frac{1}{MN_{>1}} \sum_{i=1}^{N} \sum_{m=1}^{M} c_i^{(m)} \mathbb{1}_{\{k_i^- + k_i^+ > 1\}} , \tag{S85}$$

where $\mathbb{1}_{\{\cdot\}}$ is the indicator function, $M$ is the number of samples to be drawn for each node $i$, and where the $m$-th sample, $c_i^{(m)}$, is obtained with the following procedure.

1. *Pick the hidden degrees of two neighbors*, $(\kappa_1^-, \kappa_1^+)$ and $(\kappa_2^-, \kappa_2^+)$, by sampling Eq. (S83) twice.

2. *Pick the angular distance* between node $i$ and nodes 1 and 2, $\Delta\theta_{i1}$ and $\Delta\theta_{i2}$, by sampling Eq. (S84) twice.

3. *Set the angular distance between nodes 1 and 2.* Since it is equally likely for nodes 1 and 2 to be "on the same side" or "on opposite sides" from node $i$, we set

$$\Delta\theta_{12} = \begin{cases} \min\left\{|\Delta\theta_{i1} + \Delta\theta_{i2}|, 2\pi - |\Delta\theta_{i1} + \Delta\theta_{i2}|\right\} & \text{with probability } 1/2 \\ \\ |\Delta\theta_{i1} - \Delta\theta_{i2}| & \text{with probability } 1/2 \end{cases} . \tag{S86}$$

4. *Compute the probability for nodes 1 and 2 to be connected* and set $c_i^{(m)} = P(\tilde{a}_{12} = 1|\kappa_1^-, \kappa_1^+, \kappa_2^-, \kappa_2^+, \Delta\theta_{12})$.

### E. The algorithm

The algorithm assumes that a maximal deviation tolerance, $\eta^{\text{tol}}$, has been assigned, as well as defines $\bar{c}_{\text{undir}}^{\min} = 0$, $\bar{c}_{\text{undir}}^{\max} = 1$, $\beta^{\min} = 1$ and $\beta^{\max} = 25$.

1. *Set the initial value for the parameter $\beta$* as $\beta = 1 + u$ where $u \sim \text{Uniform}(0, 1)$.

2. *Infer the hidden in/out-degrees* $\boldsymbol{\kappa} = \kappa_1^-, \kappa_1^+, \ldots, \kappa_N^-, \kappa_N^+$ by following the procedure explained in Sec. S.IV B.

3. *Infer the parameter $\nu$* by following the procedure explained in Sec. S.IV C.

4. *Estimate the triangle density $\bar{c}_{\text{undir}}$* by following the procedure explained in Sec. S.IV D.

5. *Check for convergence* by checking if $|\bar{c}_{\text{undir}} - \bar{c}_{\text{undir}}^{\text{obs}}| < \eta^{\text{tol}}$, then all $2N + 2$ parameters have been estimated within the tolerance parameters. Otherwise, proceed to step 6.

6. *Update the value of the parameter $\beta$* (bisection method):

    (a) If $\bar{c}_{\text{undir}} > \bar{c}_{\text{undir}}^{\text{obs}}$, then set $\beta^{\max} = \beta$, set $\bar{c}_{\text{undir}}^{\max} = \bar{c}_{\text{undir}}$ and proceed to step 6c.
    (b) If $\bar{c}_{\text{undir}} < \bar{c}_{\text{undir}}^{\text{obs}}$, then set $\beta^{\min} = \beta$, set $\bar{c}_{\text{undir}}^{\min} = \bar{c}_{\text{undir}}$ and proceed to step 6c.
    (c) Update $\beta$ to its new value according to

    $$\beta = \beta^{\min} + (\beta^{\max} - \beta^{\min}) \frac{\bar{c}_{\text{undir}}^{\text{obs}} - \bar{c}_{\text{undir}}^{\min}}{\bar{c}_{\text{undir}}^{\max} - \bar{c}_{\text{undir}}^{\min}} .$$

    (d) Proceed to step 2.

## S.V. TRIANGLE SPECTRA OF REAL NETWORKS

Figure S2 provides further examples of the capacity of the directed $\mathbb{S}^1$ model to reproduce the triangle spectra observed in various real directed networks.

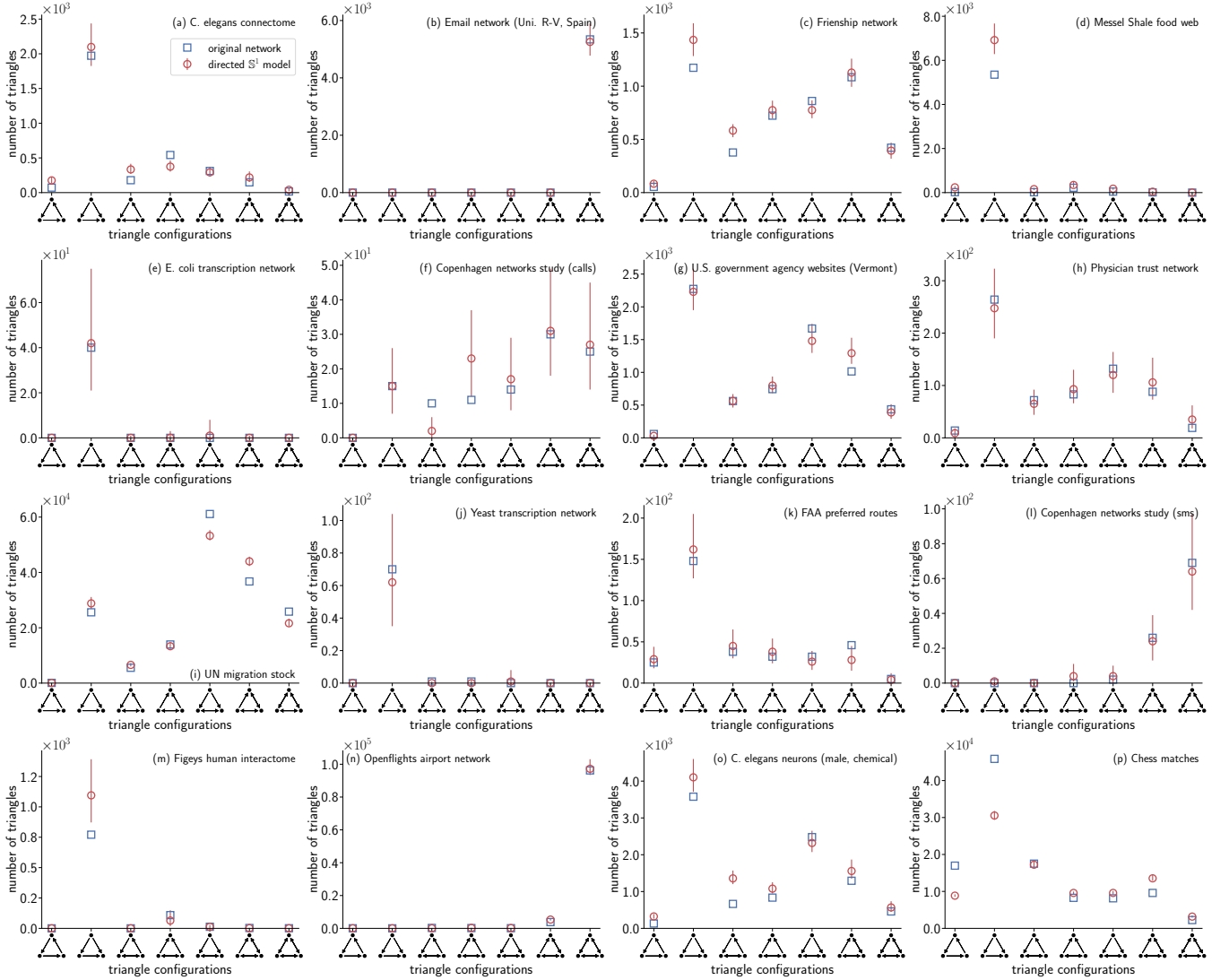

FIG. S2. **Reproducing triangle spectrum of real directed networks with the directed $\mathbb{S}^1$ model.** (a) Neural connections of the C. elegans nematode (dataset `celegansneural` [2, 3]). (b) Emails among members of a university (dataset `uni_email` [4]). (c) Friendships among high school students (dataset `add_health_comm50` [5]). (d) Messel Shale food web (dataset `messal_shale` [6]). (e) E. coli transcription network (dataset `ecoli_transcription_v1_0` [7]). (f) Social interactions among university students (dataset `copenhagen_calls` [8]). (g) Links between Vermont's government agencies websites (dataset: `us_agencies_vermont`) [9]). (h) Trust relationships among physicians (dataset `physician_trust` [10]). (i) Migration between countries (dataset `un_migrations` [11]). (j) Yeast transcription network (dataset `yeast_transcription` [12]). (k) Air traffic routes (dataset `faa_routes` [13]). (l) Social interactions among university students (dataset `copenhagen_sms` [8]). (m) Binding interactions between human proteins (dataset `interactome_figeys` [14]). (n) Regularly occurring flights among airports worldwide (dataset `openflights` [15]). (o) Networks among neurons of both the adult male and adult hermaphrodite worms C. elegans (dataset `celegans_2019_male_chemical` [16]). (p) Match outcomes between chess players (dataset `chess` [17]). Network datasets were downloaded from The Netzschleuder network catalogue and repository (https://networks.skewed.de). For each dataset, the parameters of the directed $\mathbb{S}^1$ model were adjusted using the inference procedure described in Sec. S.IV. Vertical lines show the estimated 95% confidence interval (2.5 and 97.5 percentiles).

## S.VI. USEFUL RESULTS INVOLVING THE HYPERGEOMETRIC FUNCTION

Letting $a, b \in \mathbb{C}$, $c \in \mathbb{C}\setminus\{0, -1, -2, -3, \ldots\}$ and $z \in \mathbb{Z}$, the hypergeometric function is defined by the Gauss series as [18]

$$_2F_1(a, b; c; z) = \sum_{n=0}^{\infty} \frac{\Gamma(a+n)}{\Gamma(a)} \frac{\Gamma(b+n)}{\Gamma(b)} \frac{\Gamma(c)}{\Gamma(c+n)} \frac{z^n}{n!} \tag{S87}$$

for $|z| < 1$ and elsewhere by analytic continuation. In what follows, we will use the following identity [19]

$$_2F_1(a, b; c; z) = \frac{\pi \Gamma(c)}{\sin \pi(b-a)} \left[ \frac{(-z)^{-a}}{\Gamma(c-a)\Gamma(b)\Gamma(a-b+1)} \, _2F_1\left(a, a-c+1, a-b+1; \frac{1}{z}\right) \right.$$
$$\left. - \frac{(-z)^{-b}}{\Gamma(c-b)\Gamma(a)\Gamma(b-a+1)} \, _2F_1\left(b, b-c+1, b-a+1; \frac{1}{z}\right) \right] \tag{S88}$$

valid for $\arg(1-z) < \pi$, as well as [20]

$$z \, _2F_1(a, b+1; c+1; z) = \frac{c}{b} \, _2F_1(a, b; c; z) - \frac{c}{b} \, _2F_1(a-1, b; c; z). \tag{S89}$$

We will also need Euler's reflection formula [21]

$$\Gamma(z)\Gamma(1-z) = \frac{\pi}{\sin(\pi z)} \tag{S90}$$

valid for $z \neq 0, \pm 1, \pm 2, \ldots$

We seek to evaluate the integral $\int \frac{dx}{1+x^\beta}$ for $x > 0$ and $\beta > 1$. To do so, we split the open interval $x > 0$ into two parts. First, we find for $0 < x < 1$

$$\int \frac{dx}{1+x^\beta} = \int \frac{1}{1-(-x^\beta)} dx$$
$$= \int \sum_{n=0}^{\infty} \left(-x^\beta\right)^n dx$$
$$= x \sum_{n=0}^{\infty} \frac{(-x^\beta)^n}{\beta n + 1} + C$$
$$= x \sum_{n=0}^{\infty} \frac{\Gamma(1+n)}{n!\Gamma(1)} \frac{\Gamma(\frac{1}{\beta})}{\Gamma(\frac{1}{\beta})} \frac{\Gamma(\frac{1}{\beta}+n)}{\Gamma(\frac{1}{\beta}+n)} \frac{\frac{1}{\beta}}{\frac{1}{\beta}+n} \left(-x^\beta\right)^n + C_1$$
$$= x \sum_{n=0}^{\infty} \frac{\Gamma(1+n)}{\Gamma(1)} \frac{\Gamma(\frac{1}{\beta}+n)}{\Gamma(\frac{1}{\beta})} \frac{\Gamma(1+\frac{1}{\beta})}{\Gamma(1+\frac{1}{\beta}+n)} \frac{(-x^\beta)^n}{n!} + C_1$$
$$= x \, _2F_1\left(1, \frac{1}{\beta}; 1+\frac{1}{\beta}; -x^\beta\right) + C_1 \tag{S91}$$

where $C_1 \in \mathbb{R}$. Second, we find for $x > 1$

$$\int \frac{dx}{1+x^\beta} = \int \frac{1}{x^\beta} \frac{1}{1-(-x^{-\beta})} dx$$

$$= -\int \sum_{n=0}^{\infty} \left(-x^{-\beta}\right)^{n+1} dx$$

$$= -x \sum_{n=0}^{\infty} \frac{\left(-x^{-\beta}\right)^{n+1}}{-\beta(n+1)+1} + C_2$$

$$= -x \sum_{m=1}^{\infty} \frac{\left(-x^{-\beta}\right)^m}{-\beta m + 1} + C_2$$

$$= -x \sum_{m=1}^{\infty} \frac{\Gamma(1+m)}{m!\Gamma(1)} \frac{\Gamma(-\frac{1}{\beta})}{\Gamma(-\frac{1}{\beta})} \frac{\Gamma(-\frac{1}{\beta}+m)}{\Gamma(-\frac{1}{\beta}+m)} \frac{-\frac{1}{\beta}}{-\frac{1}{\beta}+m} \left(-x^{-\beta}\right)^m + C_2$$

$$= -x \sum_{m=1}^{\infty} \frac{\Gamma(1+m)}{\Gamma(1)} \frac{\Gamma(-\frac{1}{\beta}+m)}{\Gamma(-\frac{1}{\beta})} \frac{\Gamma(1-\frac{1}{\beta})}{\Gamma(1-\frac{1}{\beta}+m)} \frac{\left(-x^{-\beta}\right)^m}{m!} + C_2$$

$$= -x \left[ {}_2F_1\left(1, -\frac{1}{\beta}; 1-\frac{1}{\beta}; -x^{-\beta}\right) - 1 \right] + C_2 \qquad (S92)$$

where $C_2 \in \mathbb{R}$. Combining Eqs. (S88) and (S90), we find

$${}_2F_1\left(1, -\frac{1}{\beta}; 1-\frac{1}{\beta}; -x^{-\beta}\right) = \frac{\frac{1}{\beta}}{1+\frac{1}{\beta}} x^\beta \, {}_2F_1\left(1, 1+\frac{1}{\beta}, 2+\frac{1}{\beta}; -x^\beta\right) + \frac{1}{\beta} \frac{\Gamma(-\frac{1}{\beta}-1)\Gamma(2+\frac{1}{\beta})}{x} . \qquad (S93)$$

Using Eq. (S89), we find

$$x^\beta \, {}_2F_1\left(1, 1+\frac{1}{\beta}, 2+\frac{1}{\beta}; -x^\beta\right) = -\frac{1+\frac{1}{\beta}}{\frac{1}{\beta}} \, {}_2F_1\left(1, \frac{1}{\beta}; 1+\frac{1}{\beta}; -x^\beta\right) + \frac{1+\frac{1}{\beta}}{\frac{1}{\beta}} \qquad (S94)$$

Combining Eqs. (S91)–(S94), we finally get

$$\int \frac{dx}{1+x^\beta} = x \, {}_2F_1\left(1, \frac{1}{\beta}; 1+\frac{1}{\beta}; -x^\beta\right) + C_3 \qquad (S95)$$

for $x > 0$ and $\beta > 1$, and where $C_3 \in \mathbb{R}$.

We also seek to evaluate the integral $\int \frac{dx}{(1+x^\beta)^2}$ for $x > 0$ and $\beta > 1$. Again, we split the open interval $x > 0$ into two parts. First, we find for $0 < x < 1$

$$\int \frac{dx}{(1+x^\beta)^2} = \int \frac{d}{d(-x^\beta)} \frac{1}{1-(-x^\beta)} dx$$

$$= \int \frac{d}{d(-x^\beta)} \sum_{n=0}^{\infty} \left(-x^\beta\right)^n dx$$

$$= \int \sum_{n=1}^{\infty} n \left(-x^\beta\right)^{n-1} dx$$

$$= x \sum_{m=0}^{\infty} \frac{(m+1)\left(-x^\beta\right)^m}{\beta m + 1} + C_4$$

$$= x \sum_{m=0}^{\infty} (m+1) \frac{\Gamma(1+m)}{m!\Gamma(1)} \frac{\Gamma(\frac{1}{\beta})}{\Gamma(\frac{1}{\beta})} \frac{\Gamma(\frac{1}{\beta}+m)}{\Gamma(\frac{1}{\beta}+m)} \frac{\frac{1}{\beta}}{\frac{1}{\beta}+m} \left(-x^\beta\right)^m + C_4$$

$$= x \sum_{m=0}^{\infty} \frac{\Gamma(2+m)}{\Gamma(2)} \frac{\Gamma(\frac{1}{\beta}+m)}{\Gamma(\frac{1}{\beta})} \frac{\Gamma(1+\frac{1}{\beta})}{\Gamma(1+\frac{1}{\beta}+m)} \frac{\left(-x^\beta\right)^m}{m!} + C_4$$

$$= x \, {}_2F_1\left(2, \frac{1}{\beta}; 1+\frac{1}{\beta}; -x^\beta\right) + C_4 \qquad (S96)$$

where $C_4 \in \mathbb{R}$. Second, we find for $x > 1$

$$\int \frac{dx}{(1+x^\beta)^2} = \int \frac{1}{x^{2\beta}} \frac{dx}{(1+x^{-\beta})^2}$$

$$= \int x^{-2\beta} \frac{d}{d(-x^\beta)} \frac{1}{1-(-x^{-\beta})} dx$$

$$= \int x^{-2\beta} \frac{d}{d(-x^{-\beta})} \sum_{n=0}^{\infty} \left(-x^{-\beta}\right)^n dx$$

$$= \int \sum_{n=1}^{\infty} n \left(-x^{-\beta}\right)^{n+1} dx$$

$$= x(-x^{-\beta}) \sum_{n=1}^{\infty} (-x^{-\beta}) \frac{d}{d(-x^{-\beta})} \frac{\left(-x^{-\beta}\right)^n}{-\beta(n+1)+1} + C_5$$

$$= x(-x^{-\beta})^2 \frac{d}{d(-x^{-\beta})} \sum_{n=1}^{\infty} \frac{\left(-x^{-\beta}\right)^n}{-\beta(n+1)+1} + C_5$$

$$= x(-x^{-\beta})^2 \frac{d}{d(-x^{-\beta})} \sum_{n=1}^{\infty} \frac{\Gamma(1+n)}{\Gamma(1)n!} \frac{-\frac{1}{\beta}}{1-\frac{1}{\beta}+n} \frac{1-\frac{1}{\beta}}{1-\frac{1}{\beta}} \frac{\Gamma(1-\frac{1}{\beta})}{\Gamma(1-\frac{1}{\beta})} \frac{\Gamma(1-\frac{1}{\beta}+n)}{\Gamma(1-\frac{1}{\beta}+n)} \left(-x^{-\beta}\right)^n + C_5$$

$$= x(-x^{-\beta})^2 \frac{-\frac{1}{\beta}}{1-\frac{1}{\beta}} \frac{d}{d(-x^{-\beta})} \sum_{n=1}^{\infty} \frac{\Gamma(1+n)}{\Gamma(1)} \frac{\Gamma(1-\frac{1}{\beta}+n)}{\Gamma(1-\frac{1}{\beta})} \frac{\Gamma(2-\frac{1}{\beta})}{\Gamma(2-\frac{1}{\beta}+n)} \frac{\left(-x^{-\beta}\right)^n}{n!} + C_5$$

$$= x(-x^{-\beta})^2 \frac{-\frac{1}{\beta}}{1-\frac{1}{\beta}} \frac{d}{d(-x^{-\beta})} \left[ {}_2F_1\left(1, 1-\frac{1}{\beta}; 2-\frac{1}{\beta}; -x^{-\beta}\right) - 1 \right] + C_5$$

$$= \frac{-\frac{1}{\beta}}{2-\frac{1}{\beta}} x(-x^{-\beta})^2 \,{}_2F_1\left(2, 2-\frac{1}{\beta}; 3-\frac{1}{\beta}; -x^{-\beta}\right) + C_5 \tag{S97}$$

where $C_5 \in \mathbb{R}$ and where we used the following identity [22] to obtain the last equality

$$\frac{d}{dz} \,{}_2F_1(a,b;c;z) = \frac{ab}{c} \,{}_2F_1(a+1, b+1; c+1; z) \,. \tag{S98}$$

Using Eqs. (S88) and (S90), Eq. (S97) becomes

$$\int \frac{dx}{(1+x^\beta)^2} = x \,{}_2F_1\left(2, \frac{1}{\beta}; 1+\frac{1}{\beta}; -x^\beta\right) + (1-\tfrac{1}{\beta})\Gamma(1-\tfrac{1}{\beta})\Gamma(1+\tfrac{1}{\beta}) + C_5 \,, \tag{S99}$$

which, combined with Eq. (S96), yields

$$\int \frac{dx}{(1+x^\beta)^2} = x \,{}_2F_1\left(2, \frac{1}{\beta}; 1+\frac{1}{\beta}; -x^\beta\right) + C_6 \,, \tag{S100}$$

for $x > 0$ and $\beta > 1$, and where $C_6 \in \mathbb{R}$.

We additionally seek to evaluate the following integral, which can be solved using Eqs. (S95) and (S100)

$$\int \frac{1}{1+x^\beta} \frac{1}{1+(\kappa x)^\beta} dx = \frac{1}{1-\kappa^\beta} \int \frac{dx}{1+x^\beta} - \frac{\kappa^\beta}{1-\kappa^\beta} \int \frac{dx}{1+(\kappa x)^\beta}$$

$$= \begin{cases} x \,{}_2F_1\left(2, \frac{1}{\beta}; 1+\frac{1}{\beta}; -x^\beta\right) + C_7 & \text{for } \kappa = 1 \\ \frac{x}{1-\kappa^\beta} \,{}_2F_1\left(1, \frac{1}{\beta}; 1+\frac{1}{\beta}; -x^\beta\right) \\ \quad - \frac{x\kappa^\beta}{1-\kappa^\beta} \,{}_2F_1\left(1, \frac{1}{\beta}; 1+\frac{1}{\beta}; -(\kappa x)^\beta\right) + C_8 & \text{for } \kappa \neq 1 \end{cases} \tag{S101}$$

with $\kappa > 0$ and $\beta > 1$, $C_7, C_8 \in \mathbb{R}$ and $x > 0$.

Letting $d \in \{1, 2\}$, we use Eq. (S88) to write

$$z \, _2F_1\left(d, \frac{1}{\beta}; 1 + \frac{1}{\beta}; -z^\beta\right) = \frac{(-1)^d \pi}{\sin(\frac{\pi}{\beta})} \frac{\Gamma(1 + \frac{1}{\beta})}{\Gamma(1 + \frac{1}{\beta} - d)} \left[\sum_{n=0}^\infty \frac{(-1)^n \Gamma(d+n)\Gamma(d - \frac{1}{\beta} + n)}{\Gamma(d - \frac{1}{\beta})\Gamma(\frac{1}{\beta}) n! \Gamma(d - \frac{1}{\beta} + 1 + s)} z^{1-(n+d)\beta} - 1\right], \quad \text{(S102)}$$

which yields

$$\lim_{z \to \infty} z \, _2F_1\left(d, \frac{1}{\beta}; 1 + \frac{1}{\beta}; -z^\beta\right) = \frac{(-1)^{d+1} \pi}{\sin(\frac{\pi}{\beta})} \frac{\Gamma(1 + \frac{1}{\beta})}{\Gamma(1 + \frac{1}{\beta} - d)}, \quad \text{(S103)}$$

and more specifically

$$\lim_{z \to \infty} z \, _2F_1\left(1, \frac{1}{\beta}; 1 + \frac{1}{\beta}; -z^\beta\right) = \frac{\pi}{\beta} \frac{1}{\sin(\frac{\pi}{\beta})} \quad \text{(S104)}$$

and

$$\lim_{z \to \infty} z \, _2F_1\left(2, \frac{1}{\beta}; 1 + \frac{1}{\beta}; -z^\beta\right) = \frac{\pi(\beta - 1)}{\beta^2} \frac{1}{\sin(\frac{\pi}{\beta})}. \quad \text{(S105)}$$


\fi

\end{document}